\documentclass{pasj01}
\draft 
\Received{$\langle$09-May-2018$\rangle$}
\Accepted{$\langle$20-June-2018$\rangle$}
\Published{$\langle$publication date$\rangle$}
\begin{document}

\title{ALMA View of the Circum-nuclear Disk of  the Galactic Center; Tidally-disrupted Molecular Clouds falling to the Galactic Center}
\author{Masato Tsuboi$^{1, 2}$, Yoshimi Kitamura$^1$, Kenta Uehara$^2$,  Takahiro Tsutsumi$^3$, Ryosuke Miyawaki$^4$,  Makoto Miyoshi$^5$, and Atsushi Miyazaki$^6$}%
\altaffiltext{1}{Institute of Space and Astronautical Science, Japan Aerospace Exploration Agency,\\
3-1-1 Yoshinodai, Chuo-ku, Sagamihara, Kanagawa 252-5210, Japan }
\email{tsuboi@vsop.isas.jaxa.jp}
\altaffiltext{2}{Department of Astronomy, The University of Tokyo, Bunkyo, Tokyo 113-0033, Japan}
\altaffiltext{3}{National Radio Astronomy Observatory,  Socorro, NM 87801-0387, USA}
\altaffiltext{4}{College of Arts and Sciences, J.F. Oberlin University, Machida, Tokyo 194-0294, Japan}
\altaffiltext{5}{National Astronomical Observatory of Japan, Mitaka, Tokyo 181-8588, Japan}
\altaffiltext{6}{Japan Space Forum, Kanda-surugadai, Chiyoda-ku,Tokyo,101-0062, Japan}

\KeyWords{accretion:accretion disks${}_1$---Galaxy: center${}_2$ --- stars: formation${}_3$ --- ISM: molecules${}_4$}

\maketitle

\begin{abstract}
We present  the high angular resolution and high sensitivity images  of the  ``Circum-Nuclear Disk (CND)" and its surrounding region  of Milky Way Galaxy in the CS $J=2-1$, SiO $v=0~J=2-1$, H$^{13}$CO$^+ J=1-0$, C$^{34}$S $J=2-1$, and CH$_3$OH $J_{K_a, K_c}=2_{1,1}-1_{1,0}A_{--}$  emission lines using ALMA. 
The CND is recognized as a torus-like molecular gas with gaps in these emission lines except for the  CH$_3$OH emission line.   The inner and outer radii of the CND are estimated to be $R_{\mathrm{in}}\sim1.5$ and $R_{\mathrm{out}}\sim2$ pc, respectively. The velocities of the rotation and radial motion  are estimated to be $V_{\mathrm{rot}}\sim115$ km s$^{-1}$   and $V_\mathrm{rad}\sim23$ km s$^{-1}$, respectively. The LTE molecular gas mass  is estimated to be $M_{\mathrm{LTE}}\sim3\times10^4 M_\odot$.
We also found some anomalous molecular clouds in the surrounding region.  
One of the molecular clouds is positionally connected to a part of the CND adjacent to the ``Western Arc". However, the cloud is seen to rotate in the opposite direction to the CND. The molecular cloud would be falling currently  from the outer region to the CND and being disrupted by the tidal shear of  Sagittarius A*(Sgr A*) because the velocity  is not yet assimilated into that of the CND.
Another molecular cloud is continuously connected to the tip of the ``Eastern Arm (EA)". The velocity of  this cloud is consistent with that of the ionized gas in the EA. These facts suggest that the molecular cloud is falling from the outer region to  the vicinity of Sgr A*, being disrupted by the tidal shear, and ionized by strong UV emission from the Central Cluster  because the impact parameter of the cloud is smaller than the first cloud. These falling clouds would play an important role in  transferring material from the outer region to the CND and/or the vicinity of Sgr A*.
\end{abstract}

\section{Introduction}
The Galactic Center is the nuclear region of the nearest spiral galaxy, Milky Way Galaxy.  Sagittarius A$^\ast$ (Sgr A$^\ast$) is a counterpart of the Galactic Center Black Hole (GCBH) in the regime from radio to X-ray, which is located very near the dynamical center of the galaxy \citep{Reid} and has a mass of $\sim4\times10^6 $M$_\odot$  (e.g. \cite{Ghez, Gillessen, Boehle}). 
The ``Circum-Nuclear Disk (CND)" has been identified conventionally as a torus-like molecular gas around Sgr A$^\ast$, which is easily observed  by high-$J$ molecular emission lines and dust continuum (e.g. \cite{Guesten, Jackson, Marr, Christopher, Montero, Martin}). The overall kinematics of the CND had been interpreted as rotation around Sgr A$^\ast$ with a velocity of $\sim100$ km s$^{-1}$. However, the recently observed kinematics of the CND has not been fully explained only by the rotation  (e.g. \cite{Jackson, Christopher, SHUKLA, Montero, Martin}).   It is yet considerably controversial how the molecular gas is supplied to the CND  (e.g.\cite{Montero, Takekawa, Hsieh}. 
Another prominent structure in the vicinity of Sgr A$^\ast$ is the ``Galactic Center Mini-spiral (GCMS)", which is a bundle of the ionized gas streams orbiting around  Sgr A$^\ast$  (e.g. \cite{Lacy1980, Ekers1983, LO1983, SerabynLacy, Serabyn1988, Lacy1991, Roberts, Scoville, Zhao2009, Zhao2010, Tsuboi2017}) and is likely feeding the GCBH.  There remains considerable controversy about the origin and life time of the GCMS. The relation between these structures is yet an open question although the GCMS is located along the inner ridge of the CND. 

The CND itself had been observed intensively by the existing millimeter interferometers and IR telescopes as mentioned above, while the outer region surrounding the CND had not been sufficiently observed because of their insufficient sensitivities  and small field of view. 
In addition, the molecular gas in the outer region would be detected better in low-$J$ molecular emission lines because of their lower excitation (e.g. \cite{Wright}) although the CND itself is prominent in high-$J$ molecular emission lines. 
The Atacama Large Millimeter/Submillimeter Array (ALMA) can detect both the CND itself and the molecular gas in the outer surrounding region because ALMA has an unprecedented high sensitivity compared to {previous} telescopes. The more detailed gas kinematics of the wider area including the CND must provide key information to  address the issues above mentioned.   Therefore, we have observed the molecular gas in the wider region using ALMA.

The distance to the Galactic Center is assumed to be 8 kpc in this paper  (e.g. \cite{Ghez, Gillessen, Schodel2009, Boehle}). Then, $1\arcsec$ corresponds to about 0.04 pc at the distance.  We use the Galactic coordinates. The directions used in this paper, for example ``north", ``south", are referred to the Galactic coordinates.
We also use the traditional nomenclature of the substructures of the GCMS because it had been established although it had been named referring to the equatorial coordinates.

\section{Observation and Data Reduction}
We have produced a 136 pointing mosaic of the 12-m array and a 68 pointing mosaic of the 7-m array (ACA) which cover a $330\arcsec \times 330\arcsec$ area including  the CND and the ``Galactic Center 50 km s$^{-1}$ Molecular Cloud (50MC)",   a most conspicuous star forming region  in the vicinity of  Sgr A$^\star$,   in the CS $J=2-1$ ($97.980953$ GHz), SiO $v=0~J=2-1$ ($86.846995$ GHz), H$^{13}$CO$^+ J=1-0$ ($86.754288$ GHz), C$^{34}$S $J=2-1$ ($96.41950$ GHz), and CH$_3$OH $J_{K_a, K_c}=2_{1,1}-1_{1,0}A_{--} (97.582808$ GHz) emission lines in ALMA Cy.1 (2012.1.00080.S, PI Tsuboi, M.). 
The CS $J=2-1$ and H$^{13}$CO$^+ J=1-0$ emission lines are dense molecular gas tracers with $n$(H$_2$)$\gtrsim 10^4$ cm$^{-3}$ and $n$(H$_2$)$\gtrsim 10^5$ cm$^{-3}$, respectively. The C$^{34}$S $J=2-1$  emission line also traces the dense gas with with $n$(H$_2$)$\gtrsim 10^4$ cm$^{-3}$, but is expected to be optically thin even in the region in contrast to the optically thick CS $J=2-1$ emission line.
 The SiO $J=2-1$ emission line is a tracer of strong C-shock  with $\Delta V\gtrsim 30$km s$^{-1}$ in molecular clouds (e.g. \cite{Gusdorf, Jim}), while the CH$_3$OH emission line is known as a tracer of weak  shocks  ($\Delta V \sim 10$ km s$^{-1}$) (e.g. \cite{Hartquist}).  
The frequency range in our observation also includes the H42$\alpha$ recombination line ($85.6884$ GHz) which is a tracer of ionized gas (e.g. \cite{Tsuboi2017}). 

The resultant maps have angular resolutions of $(2\farcs3-2\farcs5)\times (1\farcs6-1\farcs8), PA\sim-30^\circ$ using ``natural weighting" sampling on the spatial frequency ($u-v$) plane, which corresponds to $(0.09-0.10) \mathrm{pc}\times (0.06-0.07) \mathrm{pc}$ at the Galactic center distance. The synthesized beams of this observation are approximately 4 times smaller than those of previous molecular line observations  (e.g. \cite{Montero, Martin}). The original velocity resolution is $1.7$ km s$^{-1}$(488 kHz).
J0006-0623, J1517-2422, J717-3342,  J1733-1304, J1743-3058, J1744-3116 and J2148+0657 were used as phase calibrators. The flux density scale was determined using Titan, Neptune and Mars. The calibration and imaging of the data were performed by CASA \citep{McMullin}.   We made the spectral line images with $\Delta V=5$ km s$^{-1}$ of the emission lines mentioned above. The rms noise levels of the resultant maps are $\sim0.002$ Jy/beam $\times$ 5 km s$^{-1}$   or $\sim 0.35$ K km s$^{-1}$.  Because the observation has a large time span of one year and seven months, the flux uncertainty is as large as 15 \%. We will present the full data of the observation and the analysis about 50MC  in another coming paper (Uehara et al. submitted).

For comparison, we analyzed  the mosaic observation data of the CND with ALMA in the CS $J=7-6$ ($342.882857$ GHz) emission line obtained from the JVO portal of NAOJ (ALMA\#2012.1.00543.S). The data of the CS $J=7-6$ emission line have an angular resolution of $4\farcs3\times 2\farcs7, PA\sim-35^\circ$ using ``natural weighting" sampling on the $u-v$ plane, which corresponds to $0.17 \mathrm{pc}\times 0.11 \mathrm{pc}$ at the Galactic center distance. The analysis of the data were also performed by CASA.

\begin{figure}
\begin{center}
\includegraphics[width=18cm, bb=0 0  1006.82 990.59]{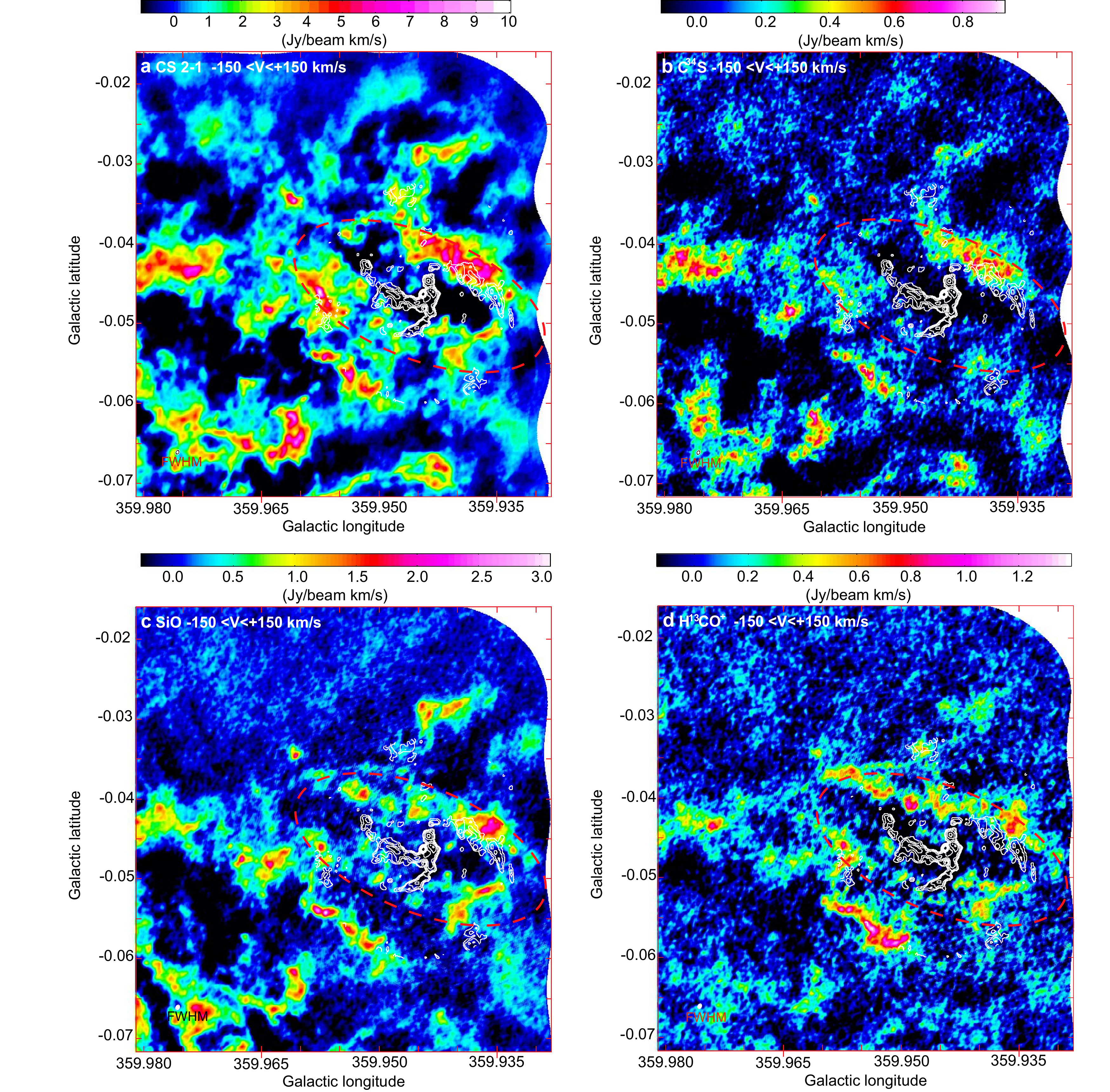}
 \end{center}
 \caption{Integrated intensity (moment 0) maps of the CND  (red dashed ellipse shows the elliptical
outline of the CND with an assumed radius $= 58 \arcsec$, PA  $= 67^\circ$, and inclination  $= 30^\circ$) and the surrounding region in  the {\bf a} CS $J=2-1$,  {\bf b} C$^{34}$S $J=2-1$,  {\bf c} SiO $v=0, J=2-1$, {\bf d} H$^{13}$CO$^{+} J=1-0$,  {\bf e} CH$_3$OH $J_{K_a, K_c}=2_{1,1}-1_{1,0}A_{--}$,  and  {\bf f} H42$\alpha$ emission lines. The integrated velocity range is $V_{\mathrm{LSR}}=-150$ to $150$ km s$^{-1}$. The FWHM beam sizes are $(2\farcs3-2\farcs5)\times (1\farcs6-1\farcs8), PA\sim-30^\circ$ shown at the lower left corners of the panels as ovals.  The contours in  each panel show the continuum emission at 100 GHz of the ``Galactic Center Mini-spiral" for comparison \citep{Tsuboi2016}.  They are set at 3.75, 7.5, 15, 30, 60, 120, 240, 480, and 960 mJy beam$^{-1}$.    {\bf g} Integrated intensity (moment 0) map in the CS $J=7-6$ emission line is shown. The data has an angular resolution of $4\farcs6\times 2\farcs7, PA\sim-35^\circ$. {\bf h} Continuum map at 670 GHz, which was obtained from the JCMT Archive. The data has an angular resolution of $8\arcsec$.}
\end{figure}
\begin{figure}
\begin{center}
\addtocounter{figure}{-1}
\includegraphics[width=18cm, bb=0 0  1025.2 991.7]{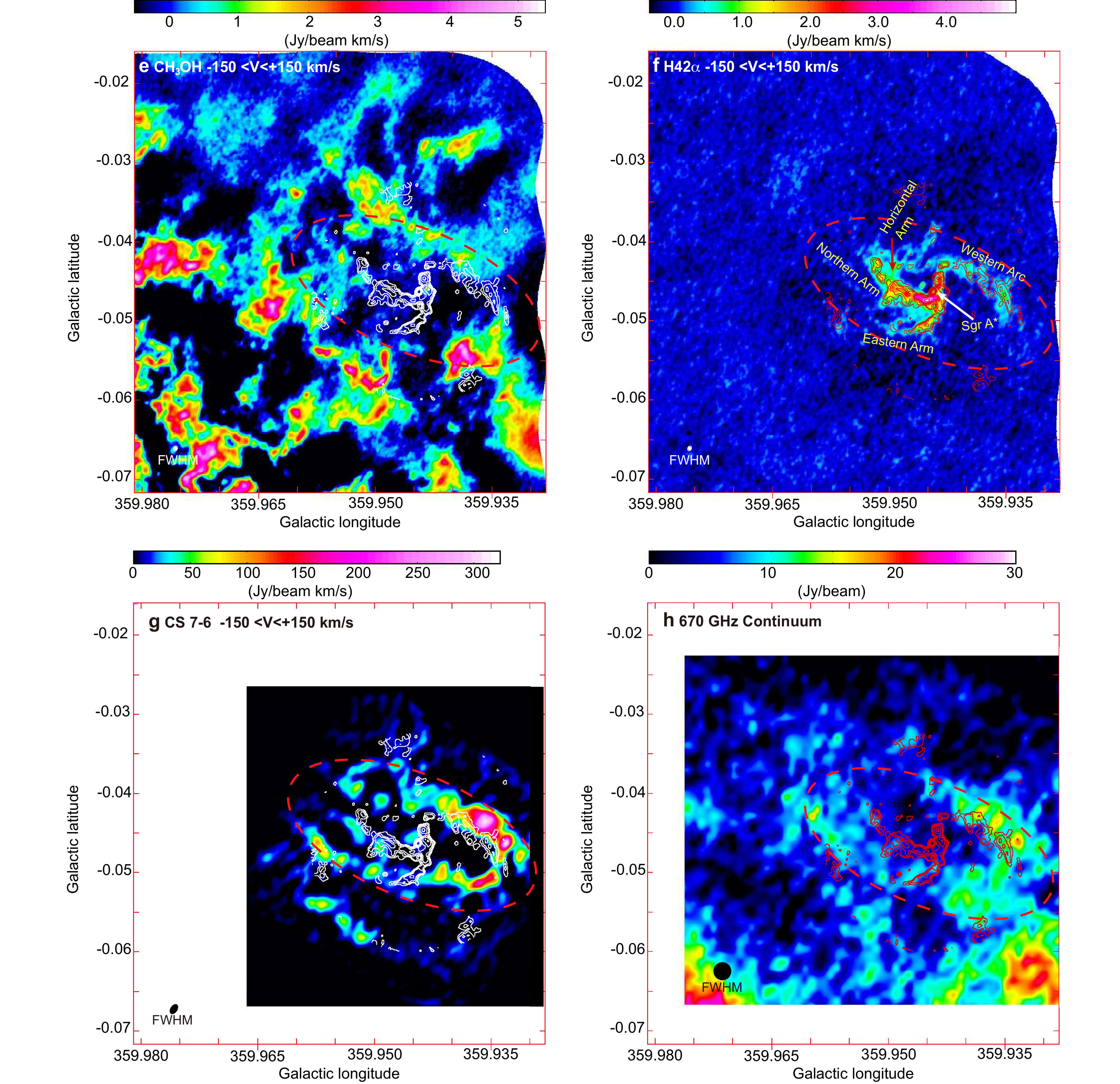}
\end{center}
\caption{Continued. }
\end{figure}
\begin{figure}
\begin{center}
\includegraphics[width=18cm,  bb=0 0  463.67 434.76]{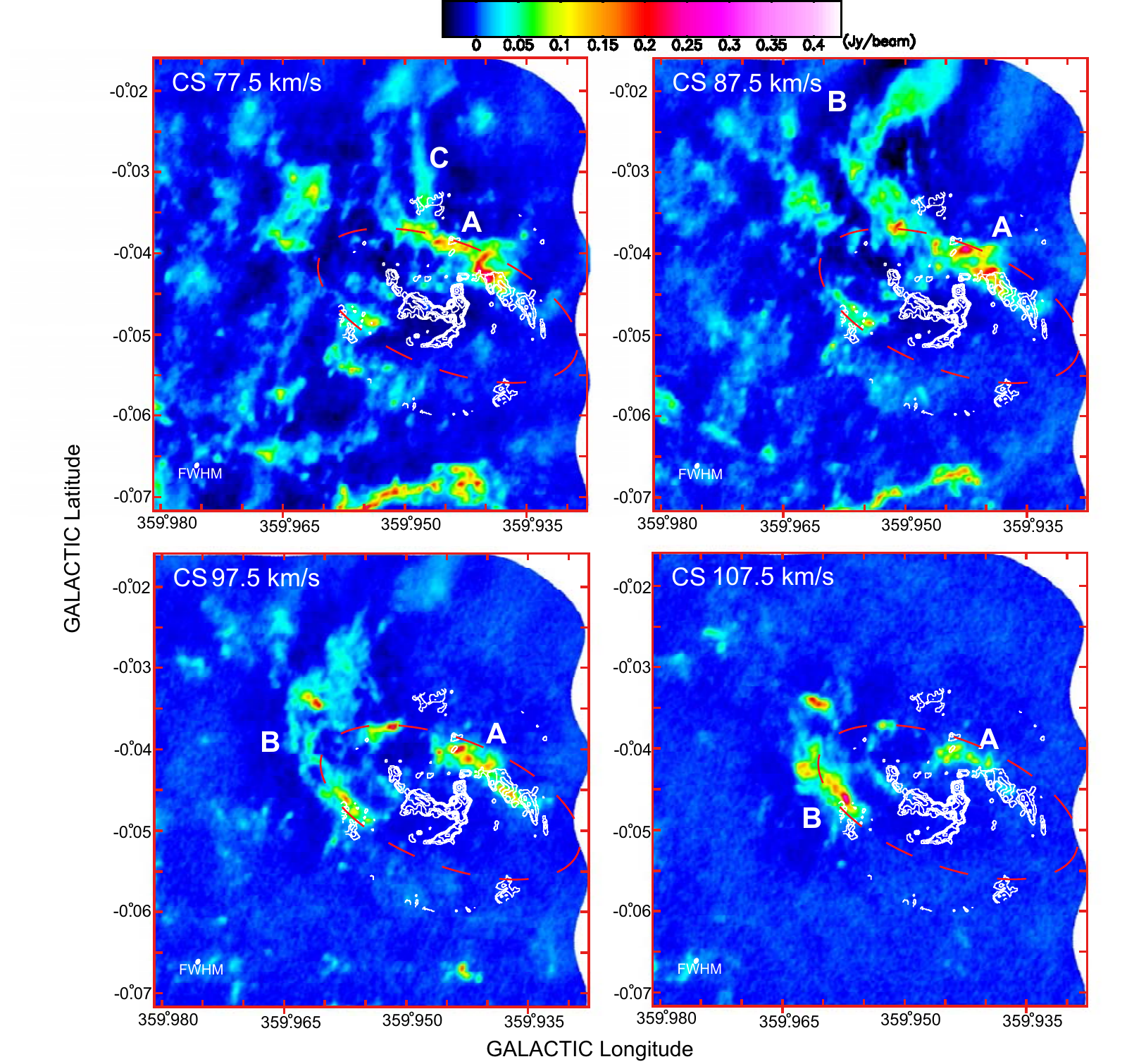}
\end{center}
\caption{Channel maps of the CND(red-dashed ellipse) and its surrounding region in the CS $J=2-1$ emission line. The central velocity range is from $V_{\mathrm{c, LSR}}=-142.5$ to $+147.5$ km s$^{-1}$, and the velocity width of each panel is $10$ km s$^{-1}$.  The angular resolution is $2\farcs3 \times 1\farcs6$ in FWHM, which is shown as the oval at the lower left corner of each panel.  The contours in the figure show the continuum emission of the ``Galactic Center Mini-spiral (GCMS)" at 100 GHz for comparison \citep{Tsuboi2016}.  They are set at 3.75, 7.5, 15, 30, 60, 120, 240, 480, and 960 mJy beam$^{-1}$.  The remaining panels are in $http://www.vsop.isas.jaxa.jp/\sim nakahara/tsuboi$. }
\end{figure}
\clearpage
\begin{figure}
\begin{center}
\includegraphics[width=18cm, bb=0 0 454.38 437.1]{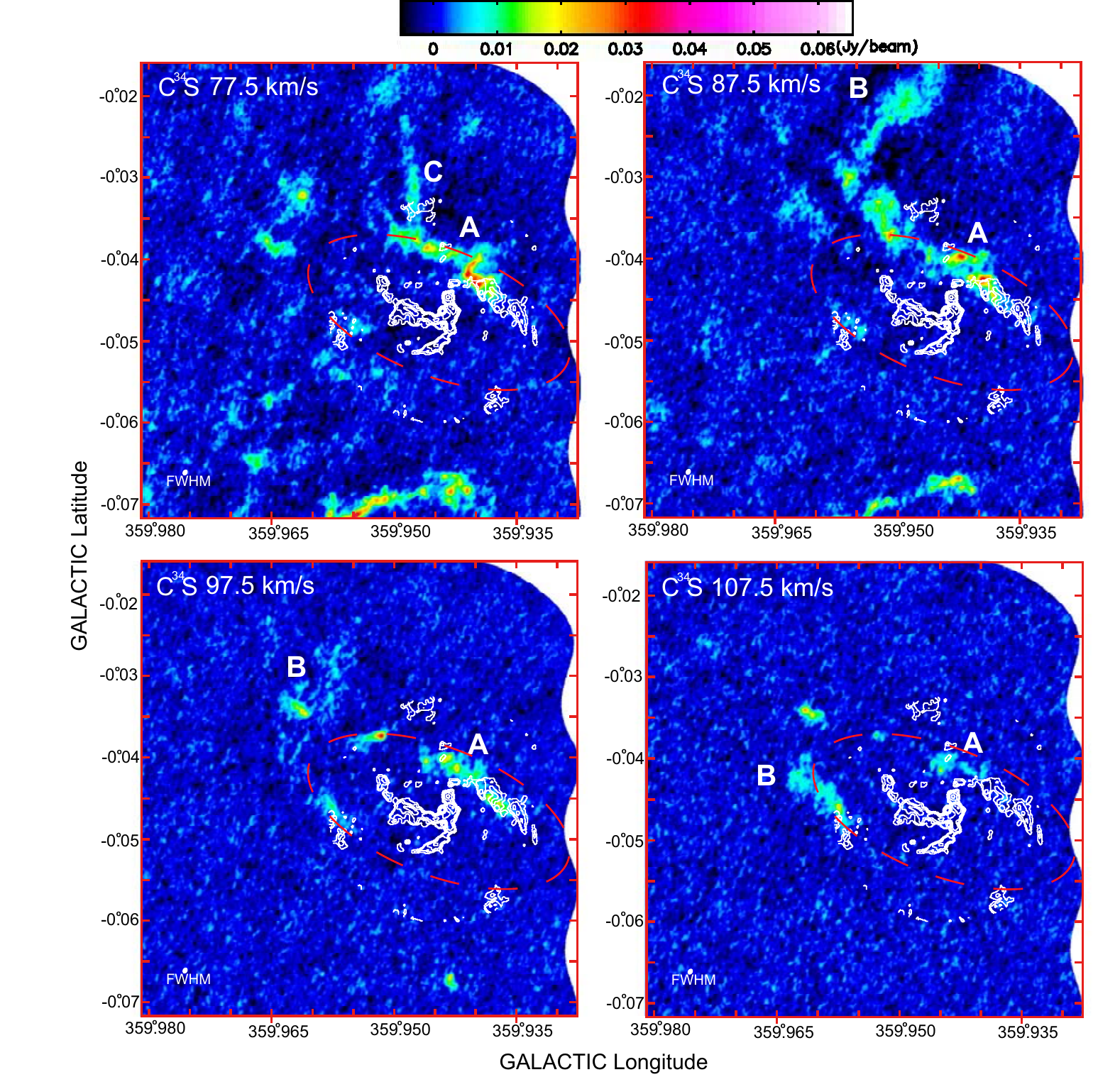}
\end{center}
\caption{Channel maps of the CND(red-dashed ellipse) and its  surrounding region in the C$^{34}$S $J=2-1$ emission line. The central velocity range is from $V_{\mathrm{c, LSR}}=-142.5$ to $+147.5$ km s$^{-1}$, and the velocity width of each panel is $10$ km s$^{-1}$.  The angular resolution is $2\farcs3 \times 1\farcs7$ in FWHM, which is shown as the oval at the lower left corner of each panel.  The contours in the figure show the continuum emission of the ``Galactic Center Mini-spiral (GCMS)" at 100 GHz for comparison \citep{Tsuboi2016}.  They are set at 3.75, 7.5, 15, 30, 60, 120, 240, 480, and 960 mJy beam$^{-1}$.  The remaining panels are in $http://www.vsop.isas.jaxa.jp/\sim nakahara/tsuboi$. } 
\end{figure}
\clearpage
\begin{figure}
\begin{center}
\includegraphics[width=18cm, bb=0 0  448.58 424.77]{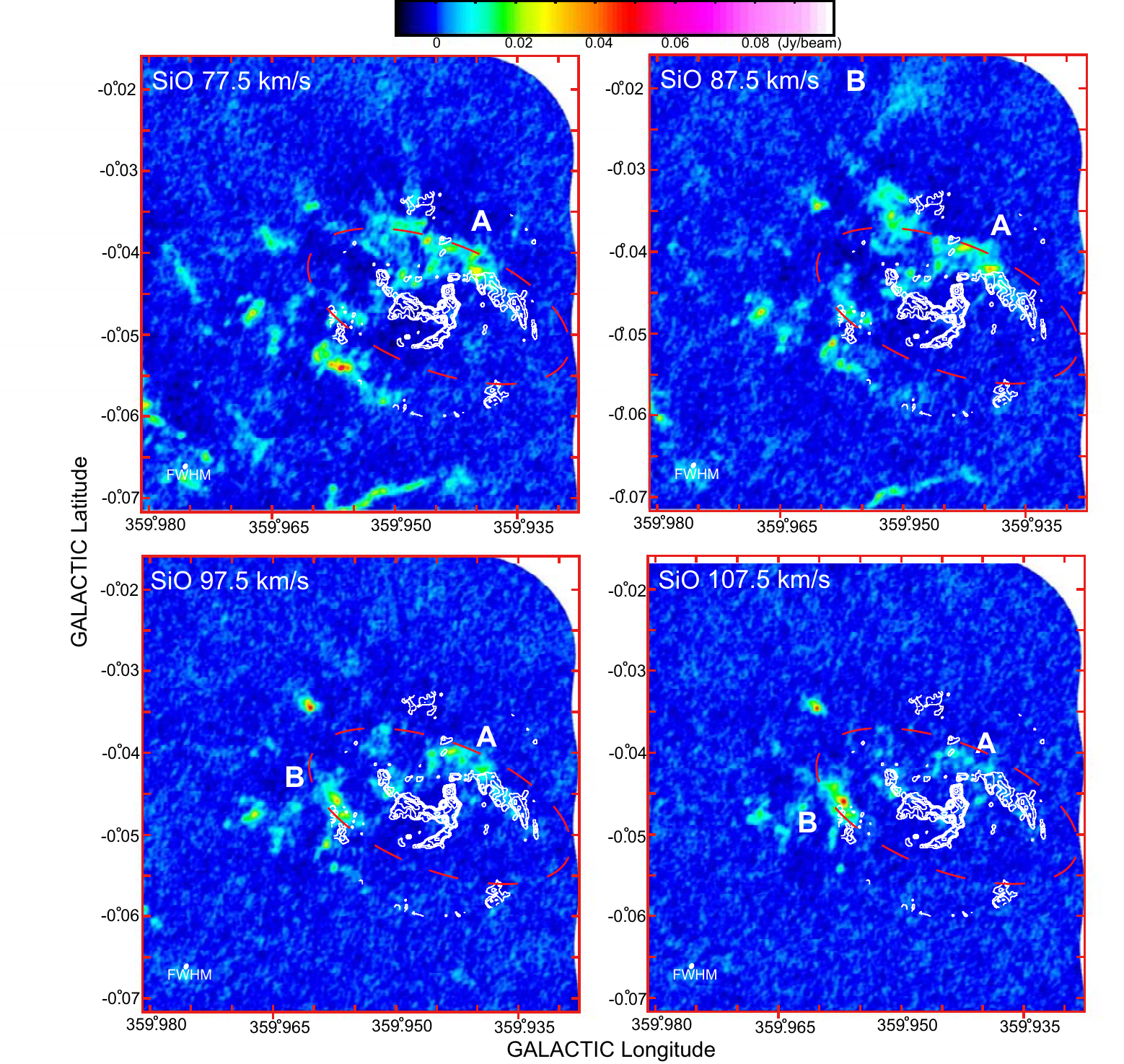}
\end{center}
\caption{Channel maps of the CND(red-dashed ellipse) and its  surrounding region in the SiO $v=0, J=2-1$ emission line. The central velocity range is from $V_{\mathrm{c, LSR}}=-142.5$ to $+147.5$ km s$^{-1}$, and the velocity width of each panel is $10$ km s$^{-1}$.  The angular resolution is $2\farcs5 \times 1\farcs8$ in FWHM, which is shown as the oval at the lower left corner of each panel.  The contours in the figure show the continuum emission of the ``Galactic Center Mini-spiral (GCMS)" at 100 GHz for comparison \citep{Tsuboi2016}.   They are set at 3.75, 7.5, 15, 30, 60, 120, 240, 480, and 960 mJy beam$^{-1}$. The remaining panels are in $http://www.vsop.isas.jaxa.jp/\sim nakahara/tsuboi$.  }  
\end{figure}
\clearpage
\begin{figure}
\begin{center}
\includegraphics[width=18cm, bb=0 0  455.26 432.68]{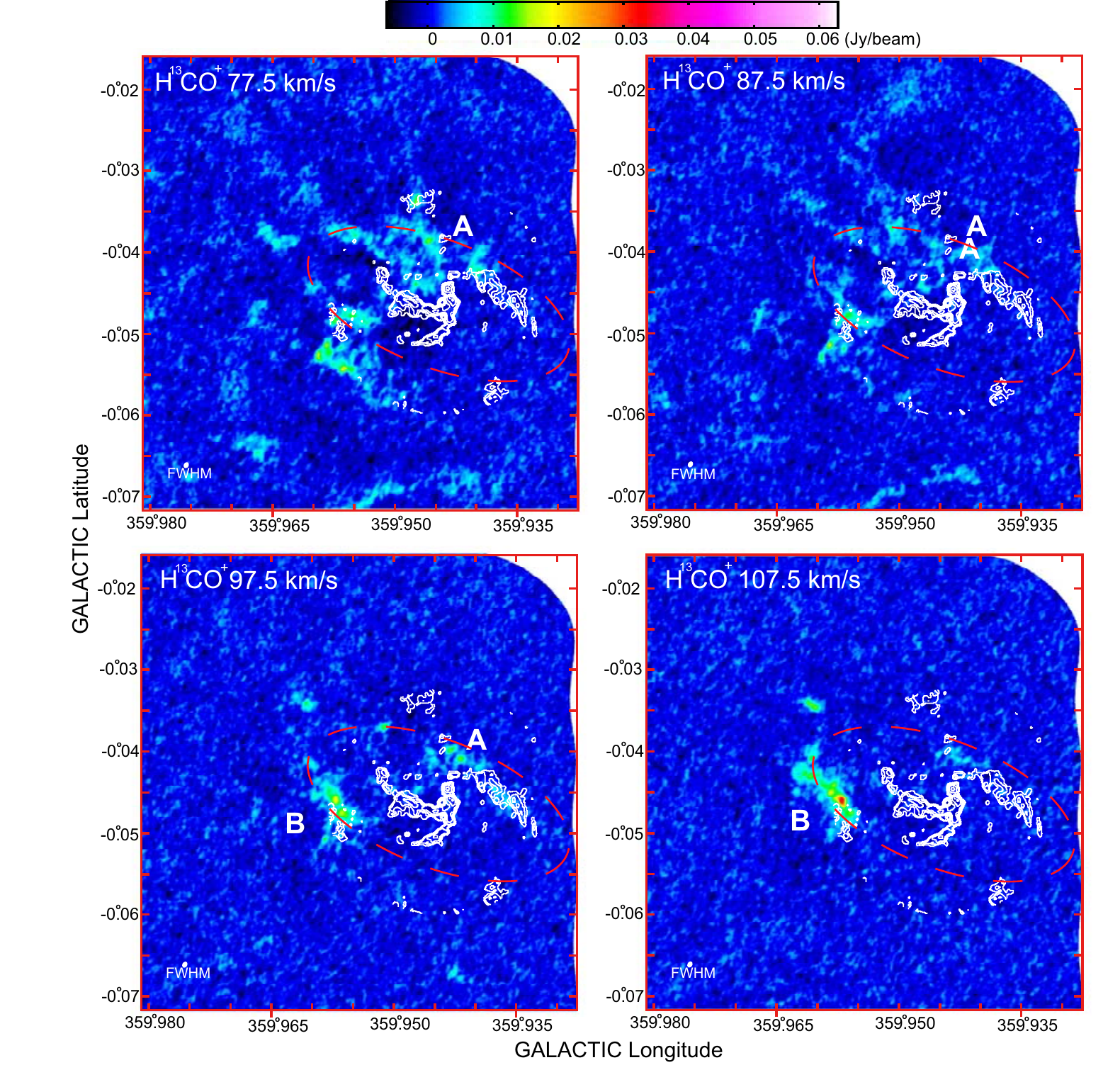}
\end{center}
\caption{Channel maps of the CND(red-dashed ellipse) and its surrounding region in the H$^{13}$CO$^+$ $J=1-0$ emission line. The central velocity range is from $V_{\mathrm{c, LSR}}=-142.5$ to $+147.5$ km s$^{-1}$, and the velocity width of each panel is $10$ km s$^{-1}$.  The angular resolution is $2\farcs5 \times 1\farcs8$ in FWHM, which is shown as the oval at the lower left corner of each panel.  The contours in the figure show the continuum emission of the ``Galactic Center Mini-spiral (GCMS)" at 100 GHz for comparison \citep{Tsuboi2016}.  They are set at 3.75, 7.5, 15, 30, 60, 120, 240, 480, and 960 mJy beam$^{-1}$.  The remaining panels are in $http://www.vsop.isas.jaxa.jp/\sim nakahara/tsuboi$.  }
\end{figure}
\clearpage
\begin{figure}
\begin{center}
\includegraphics[width=18cm, bb=0 0 454.74 432.41]{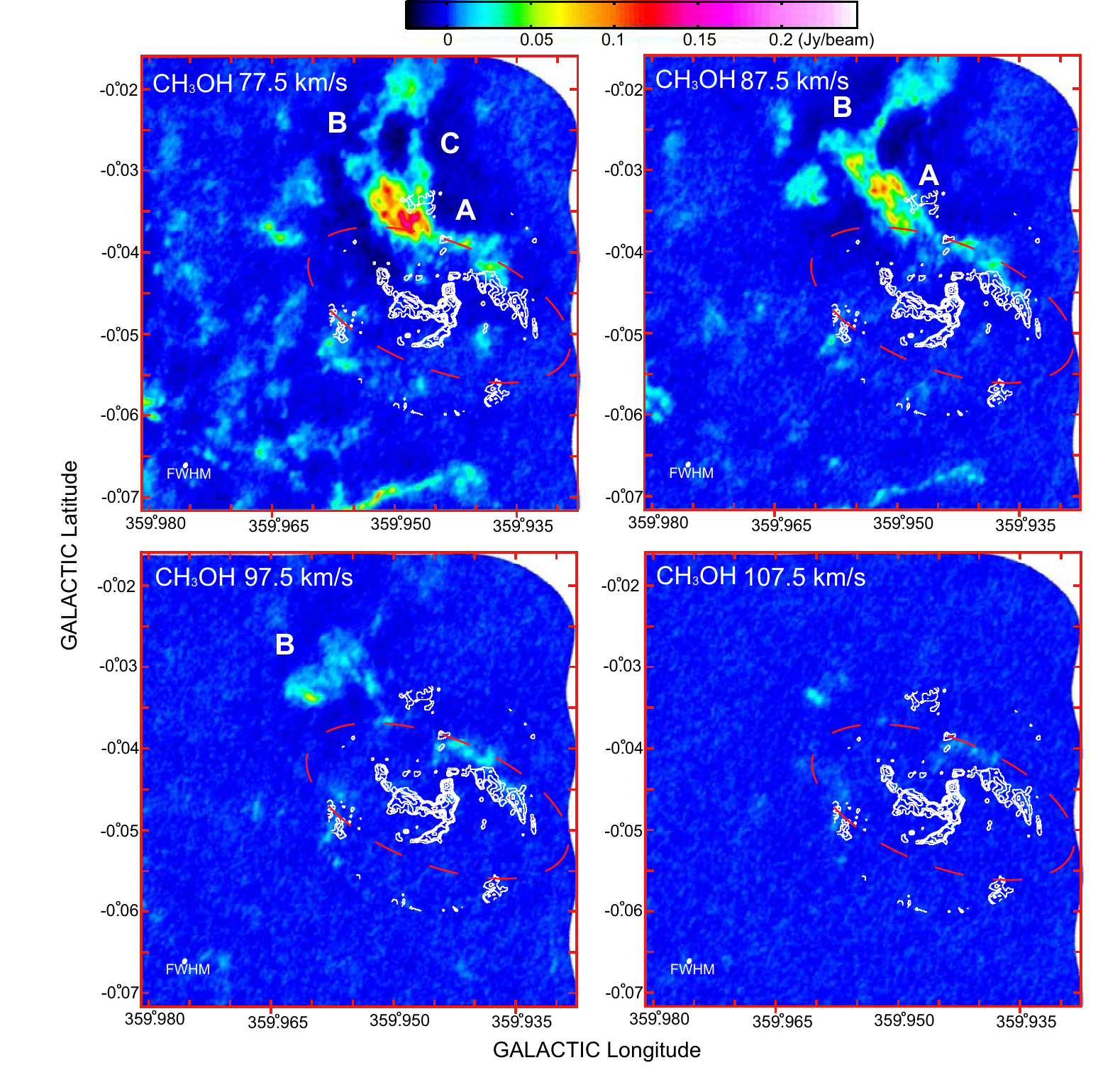}
\end{center}
\caption{Channel maps of the CND(red-dashed ellipse) and its surrounding region in the CH$_{3}$OH emission line. The central velocity range is from $V_{\mathrm{c, LSR}}=-142.5$ to $+147.5$ km s$^{-1}$, and the velocity width of each panel is $10$ km s$^{-1}$.  The angular resolution is $2\farcs3 \times 1\farcs7$ in FWHM, which is shown as the oval at the lower left corner of each panel.  The contours in the figure show the continuum emission of the ``Galactic Center Mini-spiral (GCMS)" at 100 GHz for comparison \citep{Tsuboi2016}.   They are set at 3.75, 7.5, 15, 30, 60, 120, 240, 480, and 960 mJy beam$^{-1}$.  The remaining panels are in $http://www.vsop.isas.jaxa.jp/\sim nakahara/tsuboi$.  }
\end{figure}
\clearpage
\begin{figure}
\begin{center}
\includegraphics[width=18cm, bb=0 0 455.49 441.07]{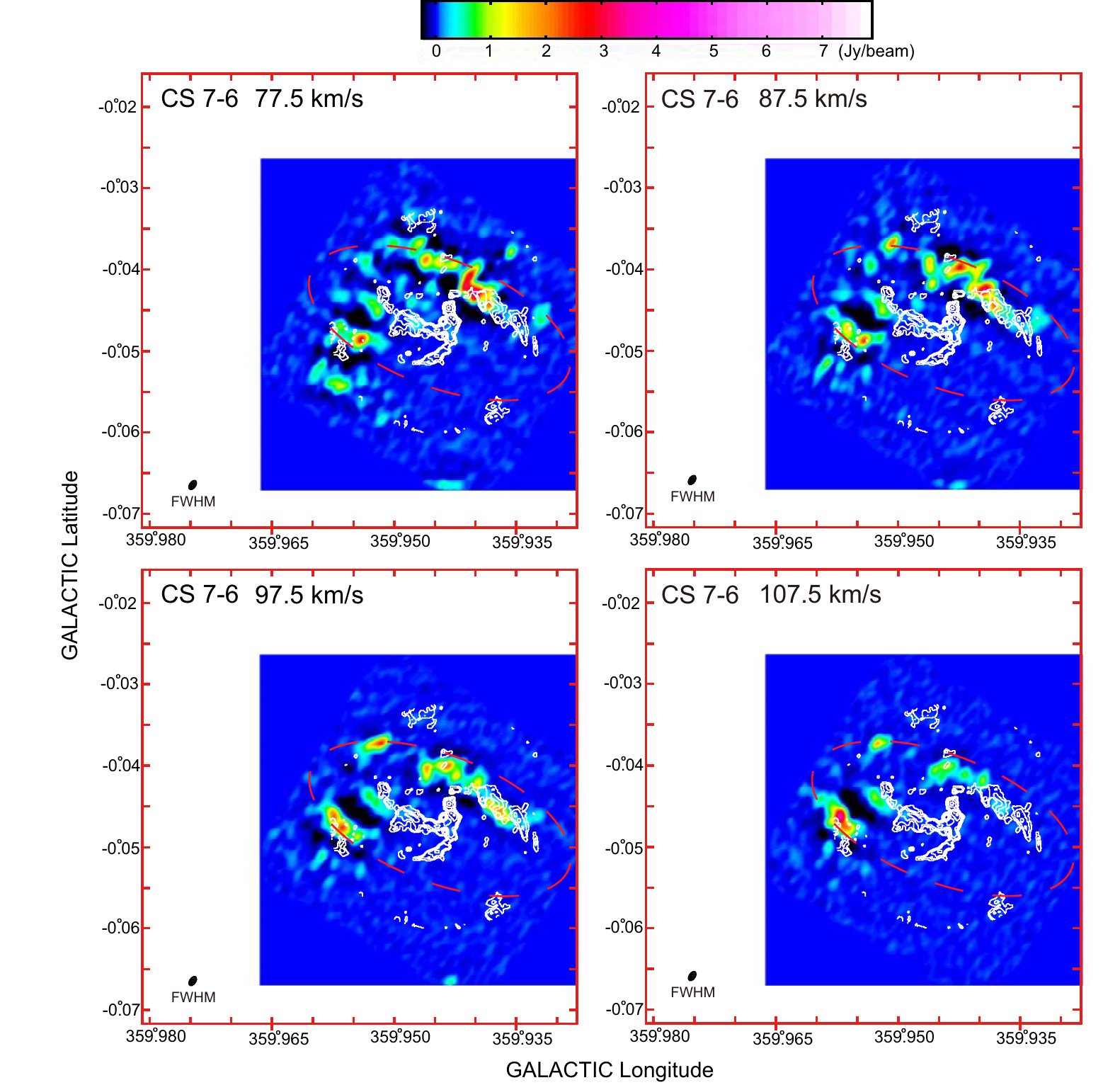}
\end{center}
\caption{Channel maps of the CND(red-dashed ellipse) and its surrounding region in the CS $J=7-6$ emission line. The central velocity range is from $V_{\mathrm{c, LSR}}=-142.5$ to $+147.5$ km s$^{-1}$, and the velocity width of each panel is $10$ km s$^{-1}$.  The angular resolution is $4\farcs6 \times 2\farcs7$ in FWHM, which is shown as the oval at the lower left corner of each panel.  The contours in the figure show the continuum emission of the ``Galactic Center Mini-spiral (GCMS)" at 100 GHz for comparison   \citep{Tsuboi2016}.   They are set at 3.75, 7.5, 15, 30, 60, 120, 240, 480, and 960 mJy beam$^{-1}$. The remaining panels are in $http://www.vsop.isas.jaxa.jp/\sim nakahara/tsuboi$. }  
\end{figure}
\clearpage

\section{Results}
\subsection{Integrated Intensity (Moment 0) Maps}
Figure 1 shows the integrated intensity (moment 0) maps of the CND(red-dashed ellipse) and its surrounding region in the CS $J=2-1$, C$^{34}$S $J=2-1$, SiO $v=0, J=2-1$,  H$^{13}$CO$^{+} J=1-0$,  and CH$_3$OH $J_{K_a, K-c}=2_{1,1}-1_{1,0}A_{--}$ and H42$\alpha$ emission lines. 
The integrated velocity range of the panels is $V_{\mathrm{LSR}}=-150$ to $150$ km s$^{-1}$. 
Because the previous observations show that the CND has the rotation velocity of $100-110$ km s$^{-1}$ and the random or radial velocity of $10-30$ km s$^{-1}$ (e.g. \cite{Guesten, Jackson, Christopher}), the integrated velocity range contains almost all of the total velocity extent of the CND.
These panels are cut from the large area mosaic data mentioned in the previous section. 
The contours in the figure show the continuum emission at 100 GHz, which mainly traces the GCMS, for comparison \citep{Tsuboi2016}.  
Figure 1 also shows the moment 0 map of the CS $J=7-6$ emission line by ALMA and the continuum map at 670 GHz by JCMT for comparison. 
Although the mosaic map of the CS $J=7-6$ emission line with a similar angular resolution had been obtained by SMA \citep{Montero}, the sensitivities of our maps are five times or more higher than those of the SMA maps.
The CND is identified as a fragmented ring surrounding the GCMS in the CS $J=2-1$, C$^{34}$S, SiO , and H$^{13}$CO$^{+} $ emission lines (Cf. \cite{Montero, Martin}), and is less prominent in the CH$_3$OH emission line.   Note that the CH3OH emission line seems to trace the surrounding gas rather than the CND itself.
There are many extended components surrounding the CND in the CS $J=2-1$, C34S, SiO, and H13CO+ lines, whereas the CS  $J=7-6$ line and 670 GHz continuum emissions mainly trace the CND. 

\subsection{Channel Maps}
Figures 2, 3, 4, 5, and 6 show the channel maps of the same region of figure 1 with the central velocities of $V_{\mathrm{c, LSR}}=-142.5$ to $+147.5$ km s$^{-1}$ in the CS, C$^{34}$S, SiO, H$^{13}$CO$^{+}$, and CH$_3$OH emission lines, respectively. The velocity width of  each panel is 10 km s$^{-1}$.  Figure 7 shows the channel maps with the same velocity range in  the CS $J=7-6$ emission line for comparison. 

Although the components of the CND identified in these channel maps are roughly consistent with those in the previous observations  (e.g. \cite{Guesten, Jackson, Marr, SHUKLA, Montero,  Martin}, significant differences are found  in the different emission lines (Cf. \cite{Martin}). 
The kinematics of the CND does not always follow a simple rotation law around Sgr A$^\ast$ as mentioned in the Introduction.  
There is a feature apparently distributed  just outside of the ``Western Arc" (WA) of the GCMS  (see also figure 1-f).  This feature is also prominent in the integrated intensity maps of several previous papers (e.g. \cite{Montero, Martin}).  However, the feature has two distinct velocity components with negative and positive radial velocities.  The positive velocity component will be discussed separately in Subsection 3.2.1.
The negative velocity component is identified in the panels in the CS emission line of $V_{\mathrm{c, LSR}}=-142.5$ to $+7.5$ km s$^{-1}$. 
The component is also identified in the C$^{34}$S, SiO, and H$^{13}$CO$^{+}$ emission lines, while the component cannot be identified in the CH$_3$OH emission line.  
The peak position of the component is seen to be shifting from west to east along the outer edge of the WA with increasing velocity.  The similar velocity shift has been observed for the WA in the H42$\alpha$ emission line (e.g. \cite{Tsuboi2017}).  Therefore this component would be physically associated with the WA and follow a simple rotation law around Sgr A$^\ast$.  The component corresponds to the ``northwestern sector" of the CND.

Another component is identified on the opposite side of the GCMS in the velocity range of $V_{\mathrm{c, LSR}}\sim-82.5$ and $+7.5$ km s$^{-1}$.  This component is seen to be corresponds to the ``southwestern sector" of the CND.
The peak position of the component would be also shifting from west to east with increasing velocity although the component has no clear association with the ionized gas of the GCMS.  The ``northwestern sector" and ``southwestern sector"  form the ``western half" of the CND, which is also identified clearly in the CS $J=7-6$ emission line.

Meanwhile, the "eastern half" of the CND has a complicated structure. There are several components with positive radial velocity seen in the eastern area of the CND. 
However, we should keep in mind that all molecular clouds seen toward the CND do not always belong to it because this is a very overlapping region as shown in Figure 1.  
The molecular clouds coming from outer regions and intruding into the CND may be molecular gas flows feeding to the CND. On the other hand, the molecular clouds passing the CND from one outer region to another outer region are thought not to belong to the CND even if they are seen toward the CND.  The distinction between the CND components and clouds overlapping the CND by chance in the same line of sight is necessary to obtain the true image of the CND.

\subsubsection{Anomaly A}
The positive velocity component apparently distributed along the WA is prominent in the panels of the CS emission line with $V_{\mathrm{c, LSR}}=+67.5$ to $+117.5$ km s$^{-1}$. The component is also identified in the C$^{34}$S and SiO emission lines, while it is faint in the  H$^{13}$CO$^{+}$  emission line. The ``southwestern half" of the component  overlaps positionally with the northwestern sector of the CND, as mentioned above.
The velocity of the component is different from that of ionized gas in the WA (e.g. \cite{Tsuboi2017}).  Thus this component  would not be physically associated with the WA.
The component  also had been identified conventionally as a part of the CND, although the positive radial velocity in the western half of the CND means that  the component should rotate oppositely around Sgr A$^\ast$ if it really belongs to the CND.  
This component is called Anomaly A (AA) hereafter. 

AA extends to the northeast in the panels of the CS emission line with $V_{\mathrm{c, LSR}}=67.5$ and $87.5$ km s$^{-1}$ (see Figure 2).   This seems to cross another component extending from southeast to northwest (Anomaly B, see below) around $l\sim359.955^\circ, b\sim-0.030^\circ$ in the panel of $V_{\mathrm{c, LSR}}=87.5$  km s$^{-1}$. 
In the CH$_3$OH emission line, the ``northeastern half" of the component with $V_{\mathrm{c, LSR}}=+77.5$ to $+87.5$ km s$^{-1} $ is prominent especially although the ``southwestern half" is faint. 
Because the CH$_3$OH emission line is known as a weak shock as mentioned in Section 2, the ``northeastern half" seems abundant in mildly shocked molecular gas  ($\Delta V \sim 10$ km s$^{-1}$).

\subsubsection{Anomaly B}
The panels with $V_{\mathrm{c, LSR}}=+87.5$ to $+117.5$ km s$^{-1}$ in the CS emission line show that  an elongated molecular cloud seems to approach SgrA$^\ast$ along an arc from the north to the outer end of the Eastern Arm (EA) of the GCMS with increasing velocity (see also figure 1-f). The component is most prominent in the CS  emission line and is identified partly in the C$^{34}$S, SiO, H$^{13}$CO$^{+}$, and CH$_3$OH emission lines.  The component is appeared to be a bundle of  filamentary substructures.  Although the component had been identified as a small protrusion from the CND since the early days of the CND observation using the low-$J$ emission lines (e.g. \cite{Guesten, Zylka, Wright}), this had never been revealed  as the bundle of  filamentary substructures by existing telescopes because it is deeply embedded in the CND.  Note that the component does not obey the simple rotation law  of the CND.  The component is called Anomaly B (AB) hereafter. The south end of AB seems to be continuously connected with the EA. On the other hand, the north end of AB is located beyond the CND.  
The  filamentary substructures remarkable in the CS emission line are faint in the SiO and H$^{13}$CO$^{+}$ emission lines except for the contact area between AB and the EA, $l\sim359.958^\circ, b\sim-0.045^\circ$.  This suggests that the strongly shocked molecular gas with $\Delta V\gtrsim 30$km s$^{-1}$ and $n$(H$_2$)$\gtrsim 10^5$ cm$^{-3}$ is abundant in the contact area.  Because the northern half of AB is prominent in the CH$_3$OH emission line, the northern half seems abundant in mildly shocked molecular gas  ($\Delta V \sim 10$ km s$^{-1}$).  

Meanwhile, the southern half of AB is not seen in the CH$_3$OH emission line. 
The CH$_3$OH molecule is reported to be destroyed easily through the cosmic-ray photodissociation reaction in the vicinity of Sgr A$^\ast$; CH$_3$OH$\to$H$_2$CO+H$_2$ \citep{Harada}. This is probably why the CND itself is faint in the CH$_3$OH emission line.  This suggests that the southern half of AB may be located near Sgr A$^\ast$  and the northern half is really farther from it.
Similarly, the CH$_3$OH maps suggest that the northeastern half of AA is farther from Sgr A$^\ast$ and the southwestern half is located near Sgr A$^\ast$.

\subsubsection{Anomaly C}
The panels with $V_{\mathrm{c, LSR}}=77.5$ km s$^{-1}$ in the CS, C$^{34}$S and CH$_3$OH emission lines show that a nearly vertical structure is located around $l\sim359.948^\circ, b\sim-0.032^\circ$ (see Figures 2, 3, and 6). This component is identified faintly in the SiO emission line (see Figure 4), but is not identified in  the H$^{13}$CO$^{+}$, and CS $J=7-6$ emission lines.  The south end of this component crosses AA.  The south end seems to be continuously connected with the Horizontal Arm of the GCMS (see Figure 1-f).  The component is called Anomaly C (AC) hereafter. 

\subsubsection{Anomaly D}
The panels with $V_{\mathrm{c, LSR}}=-72.5$ to $-32.5$ km s$^{-1}$ in the CS emission line show that  a compact molecular cloud is located around $l\sim359.948^\circ, b\sim-0.035^\circ$ (see Figure 2).  The component is prominent only in the CS  emission line. This component is identified partly in the C$^{34}$S emission line, but is faint in  the SiO, H$^{13}$CO$^{+}$, and CH$_3$OH emission lines (see Figures 3, 4, 5, and 6).
The negative radial velocity at the eastern half of the CND means that  the component should rotate oppositely around Sgr A$^\ast$ if it belongs to the CND.    Because this component is faint in the CH$_3$OH emission line, 
the cloud is likely located near Sgr A$^\ast$ as it looks. In addition, the southern edge of the component is detected faintly in the H42$\alpha$ recombination line (see Figure 1-f). The component is called Anomaly D (AD) hereafter. 

\subsubsection{-80 km s$^{-1}$ molecular cloud}
The panels with $V_{\mathrm{c, LSR}}=-102.5$ to $-42.5$ km s$^{-1}$ in the CS emission line show that  an isolated molecular cloud is located around $l\sim359.963^\circ, b\sim-0.065^\circ$. This component is also identified in the channel maps of the C$^{34}$S, SiO, H$^{13}$CO$^{+}$, and CH$_3$OH emission lines. This component has a half-shell like appearance in the panels with $V_{\mathrm{c, LSR}}=-82.5$ to $-62.5$ km s$^{-1}$ of these emission lines. As mentioned above, the prominent component in the CH$_3$OH emission line suggests that the cloud is farther from Sgr A$^\ast$.
The component would be independent from the CND. Because the intensity ratio of SiO/H$^{13}$CO$^{+}$ is $3-4$ in the component, shocked molecular gas is likely abundant in it. This component has no counterpart  in the H42$\alpha$  recombination  line (see Figure 1-f), indicating no ionized gas in it.   The component is called -80 km s$^{-1}$ molecular cloud (-80MC) hereafter. 

\subsubsection{-70 km s$^{-1}$ molecular cloud}
The panels with $V_{\mathrm{c, LSR}}=-82.5$ to $-52.5$ km s$^{-1}$ in the CS emission line show that  a molecular cloud is located around $l\sim359.969^\circ, b\sim-0.030^\circ$.  Although this component is also identified clearly in the channel maps of the C$^{34}$S emission line, it is faint in the SiO, H$^{13}$CO$^{+}$, and CH$_3$OH emission lines. This component has no counterpart  in the H42$\alpha$ recombination  line (see Figure 1-f), indicating no ionized gas in it.   The component is called -70 km s$^{-1}$ molecular cloud (-70MC) hereafter. 

\begin{figure}
\begin{center}
\includegraphics[width=18cm, bb=0 0 1012.96 950.94]{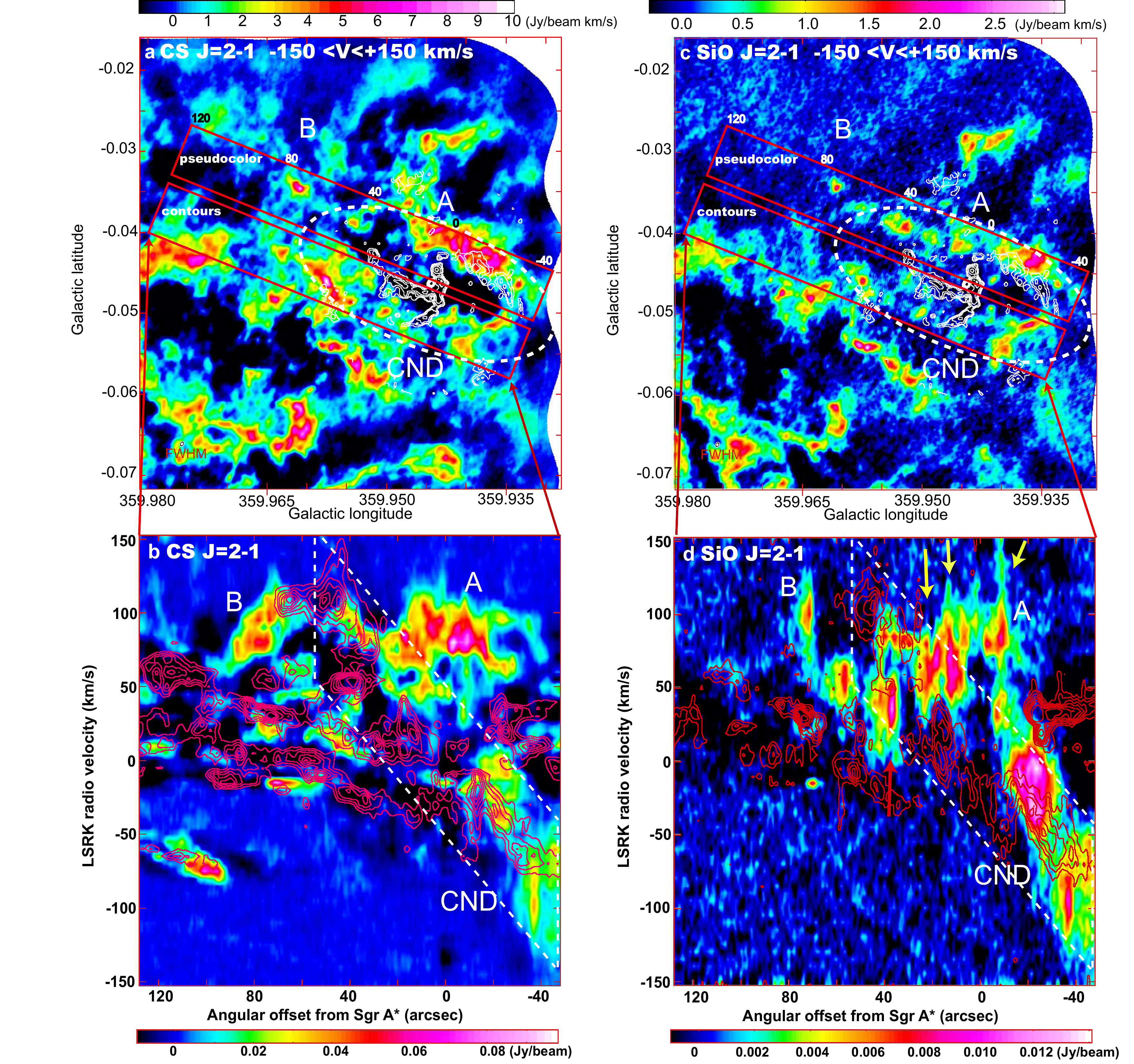}
\end{center}
\caption{ {\bf a} Moment 0 map  in the CS $J=2-1$ emission line as the finding chart of the CND(white-dashed ellipse). The contours in the  finding charts show the continuum emission of the ``Galactic Center Mini-spiral (GCMS)" at 100 GHz for comparison   \citep{Tsuboi2016}.  They are set at 3.75, 7.5, 15, 30, 60, 120, 240, 480, and 960 mJy beam$^{-1}$.  
{\bf b} Position-velocity (P-V) diagram along the major axis of the CND  in the CS $J=2-1$ emission line. The sampling areas of the diagrams are shown as the red rectangles in the finding chart. The pseudo color and contours show the components in the northern and southern areas, respectively.
 The contour levels are set at 10.3, 20.6, 30.9, 41.2, 51.5, 61.8, and 82.4 mJy beam$^{-1}$.     
The white-dashed parallelogram indicates the CND conventionally identified in the P-V diagram (e.g. \cite{Guesten, Jackson, Christopher, Montero, Martin}). ``A" and ``B" indicate the anomalies A and B of the CND shown in Figures 2, 3, 4, 5, and 6.
{\bf c} Moment 0 map  in the SiO $v=0, J=2-1$ emission line as the finding chart of the CND(white-dashed ellipse). 
{\bf d} P-V diagram along the major axis of the CND  in the SiO $v=0, J=2-1$ emission line.  The contour levels are set at 2, 4, 6, 8, 10, 12, and 16 mJy beam$^{-1}$.  
}
\end{figure}
\begin{figure}
\addtocounter{figure}{-1}
\begin{center}
\includegraphics[width=18cm, bb=0 0 1050.12 941.89]{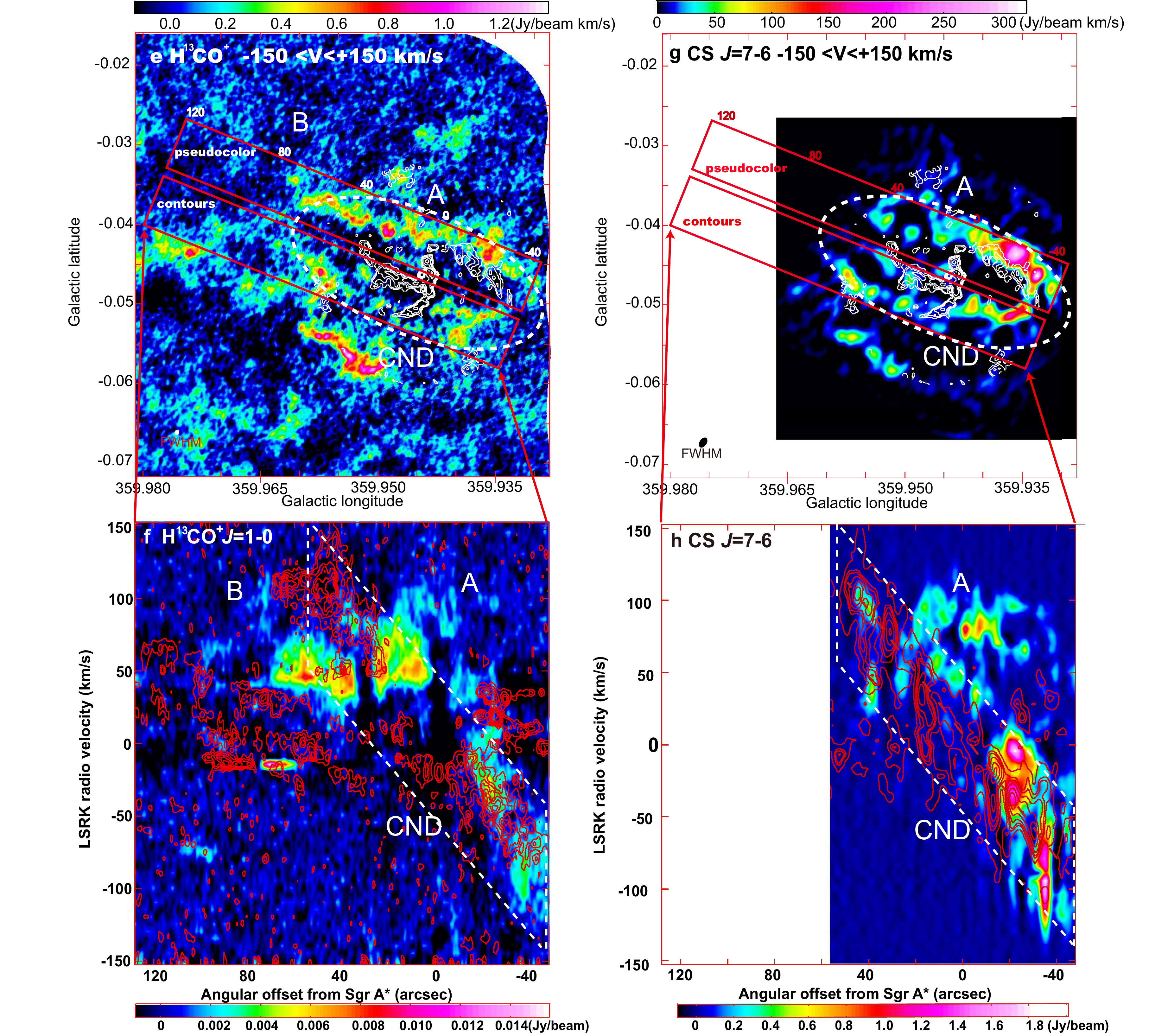}
\end{center}
\caption{Continued. {\bf e} Moment 0 map  in the H$^{13}$CO$^{+} J=1-0$ emission line as the finding chart of the CND(white-dashed ellipse). The contours in the finding charts show the continuum emission of the ``Galactic Center Mini-spiral (GCMS)" at 100 GHz for comparison   \citep{Tsuboi2016}.  They are set at 3.75, 7.5, 15, 30, 60, 120, 240, 480, and 960 mJy beam$^{-1}$.   {\bf f} P-V diagram along the major axis of the CND  in the H$^{13}$CO$^{+} J=1-0$ emission line.  The contour levels are set at 1.1, 2.2, 3.3, 4.4, 5.5, 6.6, and 8.8 mJy beam$^{-1}$. {\bf g} Moment 0 map  in the CS $J=7-6$ emission line as the finding chart of the CND(white-dashed ellipse). 
{\bf h} P-V diagram along the major axis of the CND  in the CS $ J=7-6$ emission line.  The contour levels are set at 0.2, 0.4, 0.6, 0.8, 1.0, 1.2, and 1.6 Jy beam$^{-1}$.
}
\end{figure}

\section{Discussion}
\subsection{Kinematics of the Circum-Nuclear Disk}
As mentioned in Introduction, many previous works have shown that the CND is a torus-like molecular gas with the inner boundary radius of $r_\mathrm{in}\sim1.5-1.6$ pc with a rotation velocity of $V_\mathrm{rot}\sim100-120$ km s$^{-1}$ around Sgr A$^\ast$ (e.g. \cite{Guesten, Jackson, Christopher, Montero, Martin}). 
In addition the velocity of the radial motion in the CND also has been estimated to be $\bar{V}_\mathrm{rad}\sim20-50$ km s$^{-1}$ in their works. However the observed kinematics of the CND has not been fully explained only by the rotation with radial motion,
and the outer boundary has not been clarified. 

The inner radius of the CND is estimated to be $R_{\mathrm{in}}\sim1.5$ pc from the appearances in the moment 0 maps (see Figures 8a, 8c, 8e and 8g) and the channel maps in the CS $J=2-1$, SiO $v=0, J=2-1$,  H$^{13}$CO$^{+} J=1-0$, and CS $ J=7-6$ emission lines (see Figures 2, 4, 5 and 7). 
There are many outer components which do not obey the rotation law with radial motion in the position-velocity diagrams (see Figure 8) besides the anomalies mentioned in the previous section. The outer components are distributed mainly in the velocity range of $V_{\mathrm{LSR}}\sim-30$ to $+70$ km s$^{-1}$ (see Figure 8). These components would not belong to the CND (see Figures 2, 4, 5 and 7) because they are identified to be large features outside the CND, for example, ``Northeast Arm",  ``Linear Filament"  and so on; they have been recognize as parts of the CND in the previous observations (e.g. \cite{ Montero, Martin}.  In this case, the outer radius is estimated to be as small as $R_{\mathrm{out}}\sim2$ pc. The inclination angle of the CND is estimated to be $i\sim30\pm5^{\circ}$ ($i = 0^{\circ}$ for an edge-on ring) from the aspect ratios of the appearances of the CND  in the  moment 0 maps. 

The kinematics of the CND is shown in the position-velocity diagrams along the major axis  in the CS $J=2-1$, SiO $v=0, J=2-1$,  H$^{13}$CO$^{+} J=1-0$, and CS $ J=7-6$ emission lines (see Figures 8b, 8d,  8f, and 8h). Because the CND is not clear in the CH$_3$OH emission line as mentioned in the subsection 3.1, the position-velocity diagram is not shown here.  
The CND is identified as a series of components which are roughly along an inclined line in the diagrams  (broken line parallelograms). These have been observed in the previous observations (e.g. \cite{ Montero, Martin}). 
From  the position-velocity diagrams (see Figures 8b, 8d,  8f, and 8h), the rotation velocity and radial motion of the CND are estimated to be $V_{\mathrm{rot}}\sim\frac{100}{\cos i} =115\pm10$ km s$^{-1}$   and $V_\mathrm{rad}\sim\frac{20}{\cos i}=23\pm5$ km s$^{-1}$, respectively.
The derived values are consistent with those in the previous observations (e.g. \cite{Guesten,  Marr, Montero, Martin}). 
Figure 9 shows the Keplerian orbit with the derived inner radius, outer radius, inclination angle and radial motion. The enclosed mass of $4.3\times10^6 $M$_\odot$ reproduces the observed position-velocity diagrams.

\begin{figure}
\begin{center}
\includegraphics[width=8cm, bb=0 0 408 807.06]{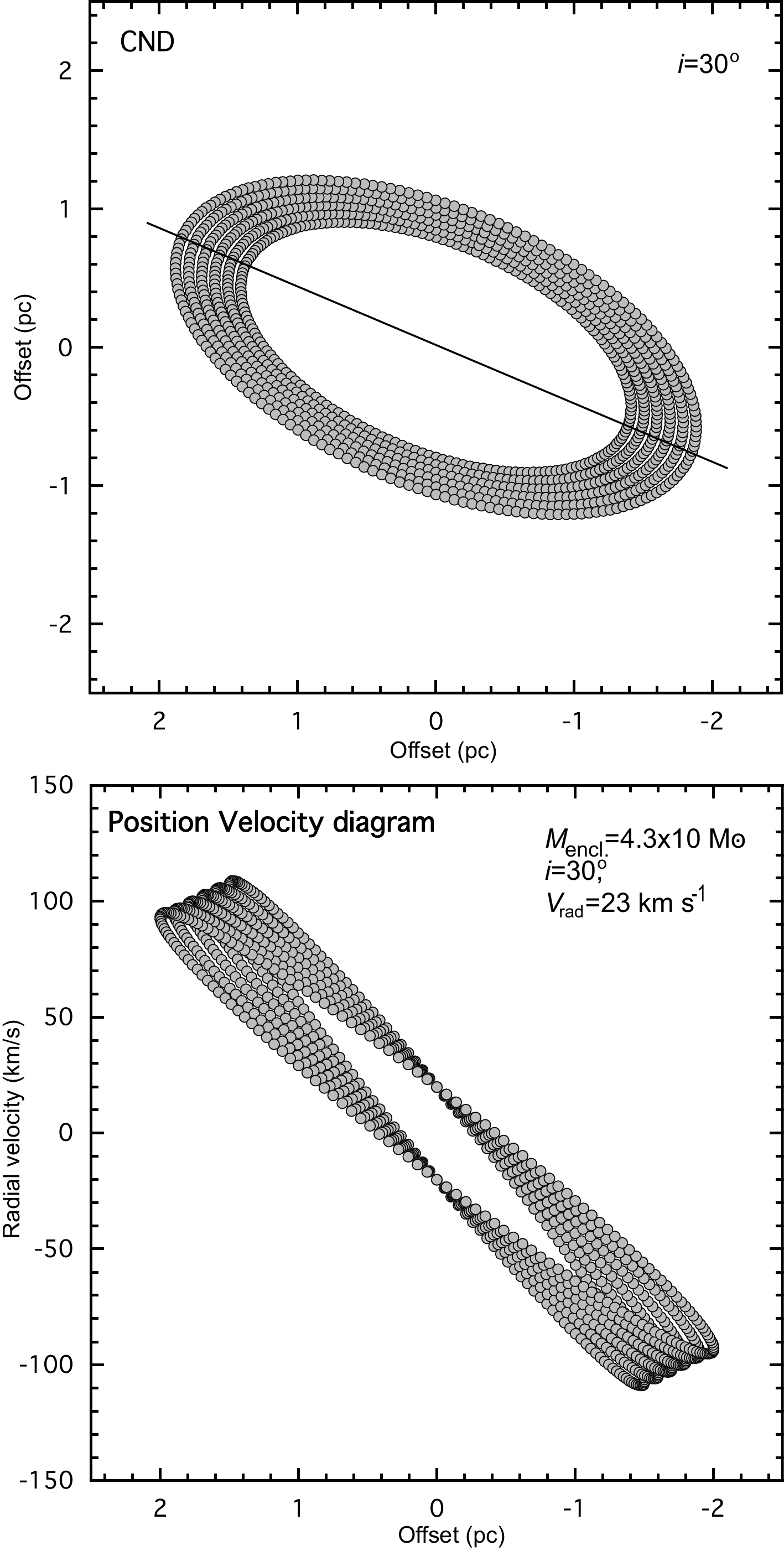}
\caption{{\bf upper panel} Trajectories of the Keplerian motions with the enclosed mass of $4.3\times 10^6$ M$_\odot$, the radius of $1.5-2.0$ pc, and the eccentricity of 0 and the radial motion of 23 km s$^{-1}$.  The inclination angle is $i=30^\circ$.   {\bf lower panel}  Position velocity diagram of the trajectories along their major axis.}
\end{center}
\end{figure}

\subsection{Physical Properties of the Circum-Nuclear Disk}
It has been reported that the moderately dense molecular gas  in the CND with $\log{N_\mathrm{H_2}}\sim4.5$ has  the kinetic temperatures of $T_{\mathrm{K}}\sim200$ K and the slightly denser gas with $\log{N_ \mathrm{H_2}}\sim5.5$ has  the kinetic temperatures of $T_{\mathrm{K}}\sim300-500$ K \citep{Requena-Torres12}.  It also has been reported that the CND would be dominated by the former component.
The kinetic temperature of the molecular gas including CS  molecules is assumed to be $T_{\mathrm{K}}=200$ K here. 

We calculate  the brightness and excitation temperatures of the CS $J=2-1$ and CS $J=7-6$ emission lines using the RADEX LVG program \citep{VanderTak} to estimate the physical properties of the CND. Figure 10a shows the relations of the CS  $J=2-1$ brightness temperature, $T_{\mathrm{B}}$(CS $J=2-1$), and the brightness temperature ratio, $T_{\mathrm{B}}$(CS $J=7-6$)/$T_{\mathrm{B}}$(CS $J=2-1$), on the plane of H$_2$ number density versus CS fractional abundance per velocity gradient are calculated at  $T_\mathrm{K}=200$ K.  Figure  10b shows the excitation temperatures of CS  $J=2-1$ and CS  $J=7-6$ emission lines, $T_{\mathrm{ex}}$(CS $J=2-1$) and $T_{\mathrm{ex}}$(CS $J=7-6$), on the same plane. The observed intensity peaks of the CS  $J=2-1$ brightness temperature in the CND and  Anomaly A are plotted in the figures as the filled and open circles, respectively. They are clearly separated in the figure. This indicates that the physical properties of CND and AA are different from each other. 

In figure 10b, the data points of the CND are distributed above the boundary where the excitation temperature of the CS  $J=2-1$ emission line is equal to the kinetic temperature. The calculated excitation temperature often has a negative sign, which nominally indicates that the emission is  a maser one. However, this is likely an artifact  made by a simplified velocity structure. Because the CS spectra in the CND do not show very high and very narrow profiles typical in the maser emission,   it is reasonable that the $J=2$ level of  the CS molecules is thermalized. Then we consider that  the excitation temperature is equal to the kinetic temperature of $T_\mathrm{K}=200$ K assumed above.
On the other hand, the excitation temperature of the CS  $J=7-6$ emission line in the CND is in the range of $T_{\mathrm{ex}}$(CS $J=7-6$)$=18-25$ K. This indicates that the $J=7$ level of the CS molecules is sub-thermally excited.
In addition, the excitation temperatures of the CS  $J=2-1$  and CS  $J=7-6$ emission lines in AA are in the ranges of $T_{\mathrm{ex}}$(CS $J=2-1$)$=56-200$ K and $T_{\mathrm{ex}}$(CS $J=7-6$)$=16-20$ K, respectively.

In figure  10a, the mean H$_2$ number density and CS fractional abundance per velocity gradient of the CND are estimated to be $n_{\mathrm{H_2}}\sim 2.2\times10^5$ cm$^{-3}$ and log$[X(\mathrm{CS})(\frac{dV}{dr})^{-1}]\sim-10.8$, respectively.  Assuming that the CS fractional abundance and the velocity width of the CS emission line  are  $X(\mathrm{CS})=10^{-8}$ and $\Delta V=50$ km s$^{-1}$, the path length of the molecular gas is estimated to be $\Delta l \sim 0.1$ pc by $X(\mathrm{CS})(\frac{dV}{dr})^{-1} \sim X(\mathrm{CS})(\frac{\Delta l}{\Delta V})$. The molecular column density based on the LVG model is derived to be $N{\mathrm{_{LVG}(H_2) }}\sim n_{\mathrm{H_2}}\times \Delta l =7\times10^{22}$ cm$^{-2}$. 
%
\begin{figure}
\begin{center}
\includegraphics[width=17cm, bb=0 0 1257.02 604.47]{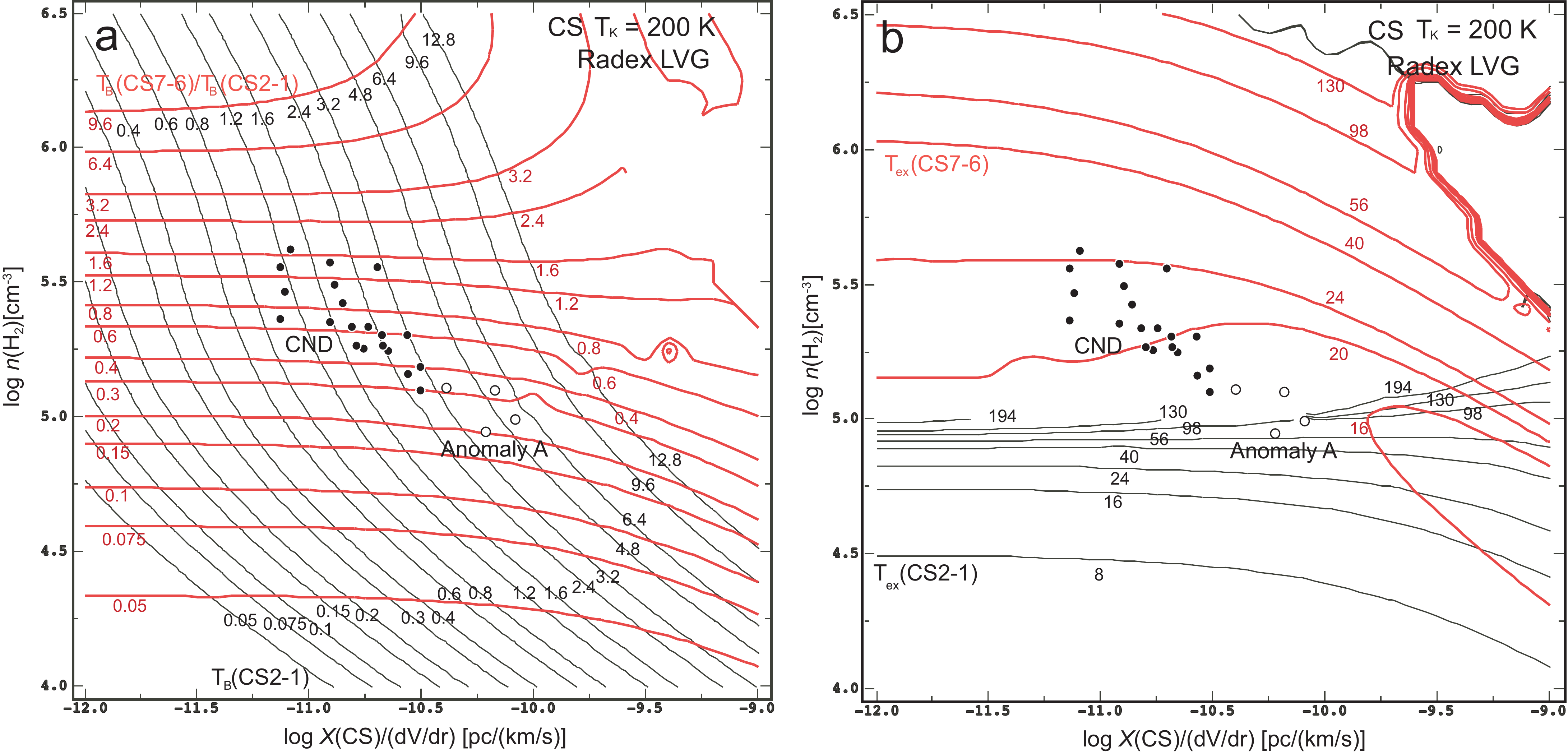}
\end{center}
\caption{\textbf{a}
Relation of  $T_{\mathrm{B}}$(CS $J=2-1$)(black contours) vs $T_{\mathrm{B}}$(CS $J=7-6$)/$T_{\mathrm{B}}$(CS $J=2-1$)(red contours) at $T_\mathrm{K}=200$ K, calculated by the RADEX LVG program  \citep{VanderTak}. 
The observed intensity peaks of the CS  $J=2-1$ brightness temperature in the the CND and  Anomaly A are plotted as filled and open circles, respectively. 
 \textbf{b}  Excitation temperatures of CS  $J=2-1$  (black contours) and CS  $J=7-6$ (red contours) emission lines  at $T_\mathrm{K}=200$ K. }
\end{figure}

 \subsection{Molecular Gas Mass of the Circum-Nuclear Disk}
The molecular gas mass of the CND  would be key information to understand the origin of the CND. 
In the previous observations, 
the molecular gas mass of the CND had been derived based on dust continuum emission maps or integrated intensity maps of molecular emission lines, because the CND is  identified easily as a ring-like feature surrounding Sgr A$^\ast$ in these maps.  However, the velocity-resolved analysis has been hard to perform  owing to  the lack of sensitivity.
The molecular gas mass of the CND even in the recent dust observations is scattered widely from $M=0.6\times10^3 $M$_\odot$ to $5\times10^4 $M$_\odot$; $0.6\times10^3 $M$_\odot$by SOFIA \citep{Lau2013}, $1-2\times10^4 $M$_\odot$ by SMA \citep{Liu2013}, $2\times10^4 $M$_\odot$ by JCMT(e.g. \cite{Oka}), and $5\times10^4 $M$_\odot$ by Hershel \citep{Etxaluze}. On the other hand, the molecular gas mass of the CND derived from molecular emission line observations 
has been not well determined
although a larger mass of $M=1-3\times10^5 {\mathrm{M}}_\odot$ has been reported \citep{Smith2013}.  Because these works have different integration areas for the CND, the various CND masses do not always mean  the ambiguity of the derivation method. It is still necessary to examine the derivation method  of the molecular gas mass and to clarify what is the boundary of the CND. The later has been done in the subsection 4.1.

 \subsubsection{CS $J=2-1$ and C$^{34}$S $J=2-1$ emission lines}
Using  the line intensities of the CS and C$^{34}$S emission lines, we derive the molecular gas mass of the CND. In that case, the ambiguity of the optical thickness of the CS emission line leads to the uncertainty of the molecular gas mass.  If the abundance ratio of CS molecule to C$^{34}$S molecule is equal to the typical cosmic ratio of the isotopes of S, $\frac{X(\mathrm{S})}{X(\mathrm{^{34}S})}=\frac{X(\mathrm{CS})}{X(\mathrm{C^{34}S})}=22.35$ $(https://physics.nist.gov/cgi-bin/Compositions/stand\_alone.pl)$, and the C$^{34}$S emission line is optically thin, the optical thickness of the CS emission line $\tau$, is estimated from the line intensity ratio of the CS and C$^{34}$S emission lines as follows; 
\begin{equation}
\label{1}
\frac{I(\mathrm{CS})}{I(\mathrm{C^{34}S})}=\frac{f(\mathrm{CS})T_\mathrm{ex}(\mathrm{CS})(1-e^{-\tau})}{f(\mathrm{C^{34}S})T_\mathrm{ex}(\mathrm{C^{34}S})(1-e^\frac{-\tau}{22.35})}.
\end{equation}
Here the integrated line intensities of CS and C$^{34}$S are $I(\mathrm{CS})=\int I_{\nu, \mathrm{CS}}(v) dv$ and $I(\mathrm{C^{34}S})=\int I_{\nu,\mathrm{C^{34}S}}(v) dv$, respectively. The components belonging to the CND are identified in the channel maps by the criteria shown in the previous section (the white-dashed ellipse  and the white-dashed parallelogram in Figure 8). The line intensities are calculated for the individual components. $f(\mathrm{CS})$ and $f(\mathrm{C^{34}S})$ are the beam filling factors of the CS and C$^{34}$S emission lines, respectively, and they are assumed to be identical here, $f(\mathrm{CS})=f(\mathrm{C^{34}S})$.
The derived optical thicknesses of the CS emission line are up to $\tau\lesssim2$ assuming that $T_\mathrm{ex}(\mathrm{CS})=T_\mathrm{ex}(\mathrm{C^{34}S})$.  Using an optical thickness correction factor, $\tau(v)/(1-e^{-\tau(v)})$, the corrected line intensity  of the CS $J=2-1$ emission line is given by
\begin{equation}
\label{2}
I\mathrm{_{corr.}(CS) }=\int I_{\nu, \mathrm{CS}}(v) \frac{\tau(v)}{1-e^{-\tau(v)}}dv.
\end{equation}

The LTE molecular column density from the CS $J=2-1$ emission line is given by 
\begin{equation}
\label{3}
N{\mathrm{_{LTE}(H_2) }}[\mathrm{cm^{-2}}]=\frac{2.35\times 10^{13} e^{\frac{2.35}{T_\mathrm{ex}}}I\mathrm{_{corr.}(CS)/\Omega\mathrm{_{beam} }}[\mathrm{Jy~beam^{-1}~km~s ^{-1}}]}
{X\mathrm{(CS)} (1-e^{\frac{-4.7}{T_\mathrm{ex}}})}.
\end{equation}
 Here the Einstein A coefficient of the CS $J=2-1$ emission line is assumed to be $A_{21}=2.2\times10^{-5}$ s$^{-1}$.
The excitation temperature of the CS $J=2-1$ emission line is estimated to be $T_{\mathrm{ex}}=200$ K  in the previous subsection using the RADEX LVG program \citep{VanderTak}.
The fractional abundance of the CS molecule is assumed to be $X\mathrm{(CS)}=\frac{N\mathrm{(CS)}}{N\mathrm{(H_2)}}=1\times10^{-8}$, which is usually used for molecular clouds in the disk region. The LTE molecular column density is approximately given by 
\begin{equation}
\label{4}
N{\mathrm{_{LTE}(H_2) }}[\mathrm{cm^{-2}}]\simeq5.1\times 10^{20} T_\mathrm{ex}I\mathrm{_{corr.}(CS)/\Omega \mathrm{_{beam} } }[\mathrm{Jy~beam^{-1}~km~s ^{-1}}].
\end{equation}
Note that the column density is proportion to $T_\mathrm{ex}$. The typical LTE molecular column density of the CND is $N{\mathrm{_{LTE}(H_2) }}\sim1\times10^{23}(T_{\mathrm{ex}}/200)$ cm$^{-2}$. 
This is consistent with the typical molecular column density based on the LVG model, $N{\mathrm{_{LTE}(H_2) }}\sim7\times10^{22}$ cm$^{-2}$, as mentioned in the previous subsection. 
 Finally, the LTE molecular gas mass is given by 
\begin{equation}
\label{5}
M_{\mathrm{LTE}}[M_\odot]=\mu[M_\odot]A\mathrm{_{beam}}[\mathrm{cm}^2]\Sigma N_{\mathrm{H_2, LTE}}=14.6(T_{\mathrm{ex}}/200)\Sigma I\mathrm{_{corr.}(CS) },
\end{equation}
where $\mu$ is  the mean molecular weight per H$_2$ molecule: $\mu = 2.8$ in amu $=2.4\times 10^{-57} M_{\odot}$, and 
$A\mathrm{_{beam}}$ is the physical area corresponding to the beam size of $\Omega \mathrm{_{beam} }$: $A\mathrm{_{beam}}=5.97 \times 10^{34} \mathrm{cm}^2 $ for the $2.3\arcsec \times 1.6\arcsec$ beam.

The total corrected integrated line intensity in the CND of the CS $J=2-1$ emission line is  $\Sigma I\mathrm{_{corr.}(CS) }=2.1\times10^3$ Jy km s$^{-1}$, whereas  that of the C$^{34}$S $J=2-1$ emission line  is 
$\Sigma I(\mathrm{C^{34}S})=8.8\times10^1$  Jy km s$^{-1}$.
The total velocity range is from $-150$ to $150$ km s$^{-1}$ in $V\mathrm{_{LSR} }$ for both emission lines. The mean line intensity ratio of the CND is calculated to be $\frac{\Sigma I\mathrm{_{corr.}(CS)}}{\Sigma I(\mathrm{C^{34}S})}\sim16.7$. Then the typical optical thickness of the CS emission line of the CND is estimated to be $\bar{\tau}\mathrm{(CND)}\sim0.63$ whereas that of AA is as large as $\bar{\tau}\mathrm{(AA)}\sim1.5$. 
The LTE molecular gas mass of the CND  is derived to be 
$M{\mathrm{_{LTE}(CND)}}=3.1\times10^4(T_{\mathrm{ex}}/200)M_\odot$.  The LTE mass  is consistent with  those derived in the previous observations mentioned above. 
In addition,  the LTE molecular gas masses of AA, AB, and AC are  derived to be $M{\mathrm{_{LTE}(AA)}}=1.3\times10^4(T_{\mathrm{ex}}/200){\mathrm{M}}_\odot$,  $M{\mathrm{_{LTE}(AB)}}=7.0\times10^3(T_{\mathrm{ex}}/200){\mathrm{M}}_\odot$, and $M{\mathrm{_{LTE}(AC)}}=1.0\times10^3(T_{\mathrm{ex}}/200){\mathrm{M}}_\odot$, respectively. However, the excitation temperature for  these components may be lower than $T_{\mathrm{ex}}=200$ K (see figure  10b). 

We also  derive the LTE molecular gas mass of the CND based on the data of the CS $J=7-6$ emission line by the same procedure for comparison.  The integrated line intensity in the CND of the CS $J=7-6$  emission line is $\Sigma I(\mathrm{CS } ~J=7-6)=4.8\times10^5$ Jy km s$^{-1}$. 
Here the Einstein A coefficient of the CS $J=7-6$ emission line is assumed to be $A_{76}=0.8\times10^{-3}$ s$^{-1}$.  
The excitation temperature of the CS $J=7-6$ emission line is calculated to be $T_{\mathrm{ex}}=18-25$ K using the RADEX LVG program \citep{VanderTak} as mentioned in the previous subsection (see figure  10b). Then the LTE molecular gas mass derived from the CS $J=7-6$ emission line becomes $M{\mathrm{_{LTE}(CND)}}= 1.5\times10^4(T_\mathrm{ex}/22) M_\odot$.

 \subsubsection{H$^{13}$CO$^{+} J=1-0$ emission line}
We derive the LTE molecular gas mass of the CND based on the observation in the H$^{13}$CO$^{+} $ emission line by the same procedure (see Figure 5).  
We consider that the $J=1$ level of the H$^{13}$CO$^{+}$ molecules is also thermalized and the excitation temperature is equal to the kinetic temperature of $T_{\mathrm{K} }= 200$ K. The Einstein A coefficient of the H$^{13}$CO$^{+} $ emission line is assumed to be $A_{10}=3.9\times10^{-5}$ s$^{-1}$.

The integrated line intensity in the CND of the H$^{13}$CO$^{+} $ emission line is $\Sigma I(\mathrm{H^{13}CO^{+}}) =9.2\times10^1$ Jy km s$^{-1}$. Then the LTE molecular gas mass of the CND from the H$^{13}$CO$^{+} $ emission line observation is derived to be $M_{\mathrm{LTE}}=1.5\times10^4(T_\mathrm{ex}/200) M_\odot$ assuming the fractional abundance of the H$^{13}$CO$^{+} $ molecule of $X\mathrm{(H^{13}CO^{+} )}=\frac{N\mathrm{(H^{13}CO^{+})}}{N\mathrm{(H_2)}}=4\times10^{-10}$ (for the CND; \cite{Amo}). 
Although the fractional abundance for the CND  is nine times larger than that for the Orion A molecular cloud, $X\mathrm{(H^{13}CO^{+} )}=5\times10^{-11}$ \citep{Ikeda},  the consistency in the derived gas masses from the CS and H$^{13}$CO$^{+}$  observations suggests that the estimations of the excitation temperature and the fractional abundance are good for the CND. 

 The LTE masses of the CND derived from the CS $J=7-6$ and H$^{13}$CO$^{+} J=1-0$ emission lines,  $M_{\mathrm{LTE}}\sim1.5\times10^4(T_{\mathrm{ex}}/200) M_{\odot}$, are half of that derived from the CS $J=2-1$ emission line, $M_{\mathrm{LTE}}\sim3.1\times10^4(T_{\mathrm{ex}}/200) M_{\odot}$.  
These relations are consistent with that the effective critical densities of the CS $J=7-6$ and H$^{13}$CO$^{+} J=1-0$ emission lines, $n{\mathrm{_{eff}}}\sim10^5$ cm$^{-3}$, are higher than that of the CS $J=2-1$ emission line, $n{\mathrm{_{eff}}}\sim10^3$ cm$^{-3}$. The derived masses of tidally disrupted molecular clouds which will be discussed in the next subsection are also summarized in Table 1. 

\begin{table}
  \caption{Physical Paramters of the Circum-Nuclear Disk and Tidally-disrupted Molecular Clouds }\label{tab:first}
   \begin{center}
    \begin{tabular}{ccccc}
    \hline
Object &$\tau$(CS2-1) &$M_{\mathrm{LTE}}$(CS2-1)$ ^1$ & $M_{\mathrm{LTE}}$(H$^{13}$CO$^+$)$^2$ & $M_{\mathrm{LTE}}$(CS7-6)$^3$  \\
    & & [$M_{\odot}$] & [$M_{\odot}$] & [$M_{\odot}$] \\
\hline
     CND & $\sim0.63$ &$3.1\times10^4(T_{\mathrm{ex}}/200)$&$1.5\times10^4(T_\mathrm{ex}/200)$&$1.5\times10^4(T_\mathrm{ex}/22)$  \\
     Anomaly A & $\sim1.5$ & $1.3\times10^4(T_{\mathrm{ex}}/200)$ &$2.3\times10^3(T_{\mathrm{ex}}/200)$ &$-$ \\
     Anomaly B & $\sim0.18$ & $7.0\times10^3(T_{\mathrm{ex}}/200)$ &$2.7\times10^3(T_{\mathrm{ex}}/200)$  &$-$ \\
     Anomaly C & $\sim3.2$ & $1.0\times10^3(T_{\mathrm{ex}}/200)$&$- $&$-$ \\     
          \hline
    \end{tabular}
  \end{center}
     $^1$ We assumed $A_{21}=2.2\times10^{-5}$ s$^{-1}$ and $X(\mathrm{CS})=1\times10^{-8}$. \\
     $^2$ We assumed  $A_{10}=3.9\times10^{-5}$ s$^{-1}$ and $X\mathrm{(H^{13}CO^{+} )}=4\times10^{-10}$ (\cite{Amo}).\\
     $^3$ We assumed  $A_{76}=0.8\times10^{-3}$ s$^{-1}$ and $X(\mathrm{CS})=1\times10^{-8}$. 
\end{table}

 \subsection{Tidally-disrupted Molecular Clouds falling to the Galactic Center}  
AA, AB, and AC  could be clouds falling from the outer regions onto the CND or the GCMS because they apparently connect with the inner structures.
The channel maps with $V_{\mathrm{c, LSR}}=77.5$  to $117.5$ km s$^{-1}$ in the CS emission line show that  AA has a continuous appearance from the northeast to the vicinity of WA (see Figure 2). 
Although the northeastern part of AA is prominent in the CH$_3$OH emission line,  the inner part apparently adjacent to WA is faint in the emission line  (see the channel maps with $V_{\mathrm{c, LSR}}=77.5$  to $97.5$ km s$^{-1}$  in the CH$_3$OH emission line in Figure 6).  On the other hand,  although the northeastern part is faint in the SiO emission line, the inner part is prominent in the emission line (see Figure 4). The part is also identified by SOFIA as a compact component in extremely high-$J$ CO emission lines (up to $J=16-15$) \citep{Requena-Torres12}. 
Because the CH$_3$OH emission line is activated by shock with milder shock velocity, $\Delta V \sim 10$ km s$^{-1}$, than that of the SiO emission line, the northeastern part is abundant in mildly shocked molecular gas. The SiO enhancement shows that the inner part  is abundant in strongly shocked molecular gas. It is likely that the CH$_3$OH emission line of the inner part is also enhanced by the shock,but the observed weak line suggests the depletion of the CH$_3$OH molecules in the inner part. 

AA has wide velocity and narrow sub-features in the PV diagram of the SiO emission line (arrows in figure  8). These are also identified in the channel maps  of the SiO emission line (see figure  4).
 These  properties suggest  that there are several spots including abundant strongly shocked molecular gas in AA.
Because there is no counterpart of the ionized gas, these would be activated by C-type shocks. 
The FWHM velocity widths of the components in the SiO emission line are up to $\Delta V_{\mathrm{FWHM}}=50$ km s$^{-1}$. 
 This may be interpreted as  the shock velocity between the approaching gas and the CND, $V_{\mathrm{shock}}\sim50$ km s$^{-1}$.
The magnetic field strength around the CND is as strong as $B\sim 4$ mG, which was estimated by the Zeeman effect of the OH maser line  \citep{Yusef-Zadeh1996}.  While the Alfv\'en velocity in molecular material is given by 
\begin{equation}
\label{1}
V_\mathrm{A}\mathrm{[km~s^{-1}]}\sim1300\times B\mathrm{[mG]}/\sqrt{n\mathrm{(H_2)}\mathrm{[cm^{-3}]}},
\end{equation}
when the molecules are assumed to be frozen in the magnetic field. Based on the LVG model,  the number density of AA  is as high as $\mathrm{n(H_2)}\sim0.6-1.2\times10^5$ cm$^{-3}$ (see figure 10b).  The Alfv\'en velocity is estimated to be $V_\mathrm{A}\sim15-20$ km s$^{-1}$. 
Because the Alfv\'en velocity is as large as the shock velocity, it is plausible that the shock wave becomes C-type.

The CH$_3$OH molecule is reported to be destroyed easily through the cosmic-ray photodissociation reaction in the vicinity of Sgr A$^\ast$; CH$_3$OH+c.r.$\to$H$_2$CO+H$_2$ \citep{Harada}. The disappearance of AA in the CH$_3$OH emission line  suggests that the inner part of AA is really located near Sgr A$^\ast$ rather than chance coincident.  We argue that AA is falling  from the outer region to the CND, being disrupted by the tidal shear of Sgr A$^\star$ and affected by cosmic-ray.  AA is probably a route of feeding molecular gas to the CND. If this is the case, it is possible that the falling cloud does not obey the rotation law of the CND as observed because the orbit depends on the initial conditions including impact parameter and the conditions  would be various. The molecular cloud would be being fallen currently  from the outer region to the CND because  the dynamical relaxation  by collisions with molecular gas in the CND has not yet occurred. 
In addition, the photodissociation by cosmic-ray is probably why it is hard to depict the CND itself in the CH$_3$OH emission line (see Figure 1-e). 

\begin{figure}
\begin{center}
\includegraphics[width=18cm, bb=0 0 734.15 414.94]{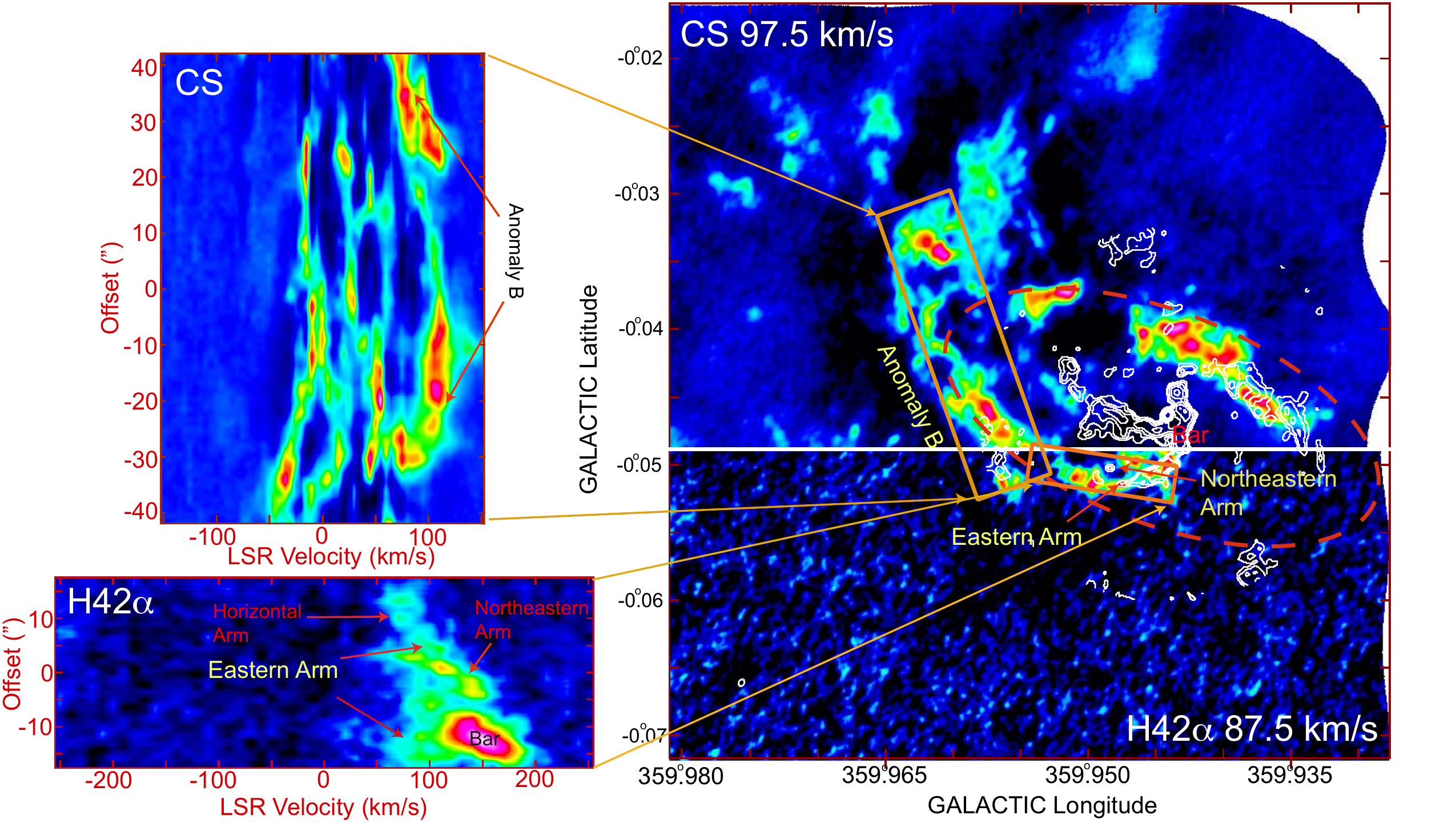}
\end{center}
\caption{Right panel: Integrated intensity map of Anomaly B at $V_{\mathrm{LSR}}=97.5$ km s$^{-1}$ in the CS $J=2-1$ emission line (upper part) and Easter Arm at $V_{\mathrm{LSR}}=87.5$ km s$^{-1}$ in the  H42$\alpha$ recombination line (lower part). The central velocities are shown at the upper left and lower right corners of the upper and lower parts, respectively. The velocity widths of the two parts are 10 km s$^{-1}$.   The red dashed ellipse shows the elliptical outline of the CND.  The white contours show the continuum emission at 100 GHz of the ``Galactic Center Mini-spiral" at 100 GHz for comparison \citep{Tsuboi2016}.  
Upper left panel: PV diagram along Anomaly B in the CS $J = 2 - 1$ emission line. The sampling area is shown by the upper red rectangle in the right panel.
Lower left panel: PV diagram along Eastern Arm in the H42$\alpha$ recombination line. The sampling area is shown by the lower red rectangle in the right panel.}
\end{figure}
\begin{figure}
\begin{center}
\includegraphics[width=18cm, bb=0 0 921.94 415.08]{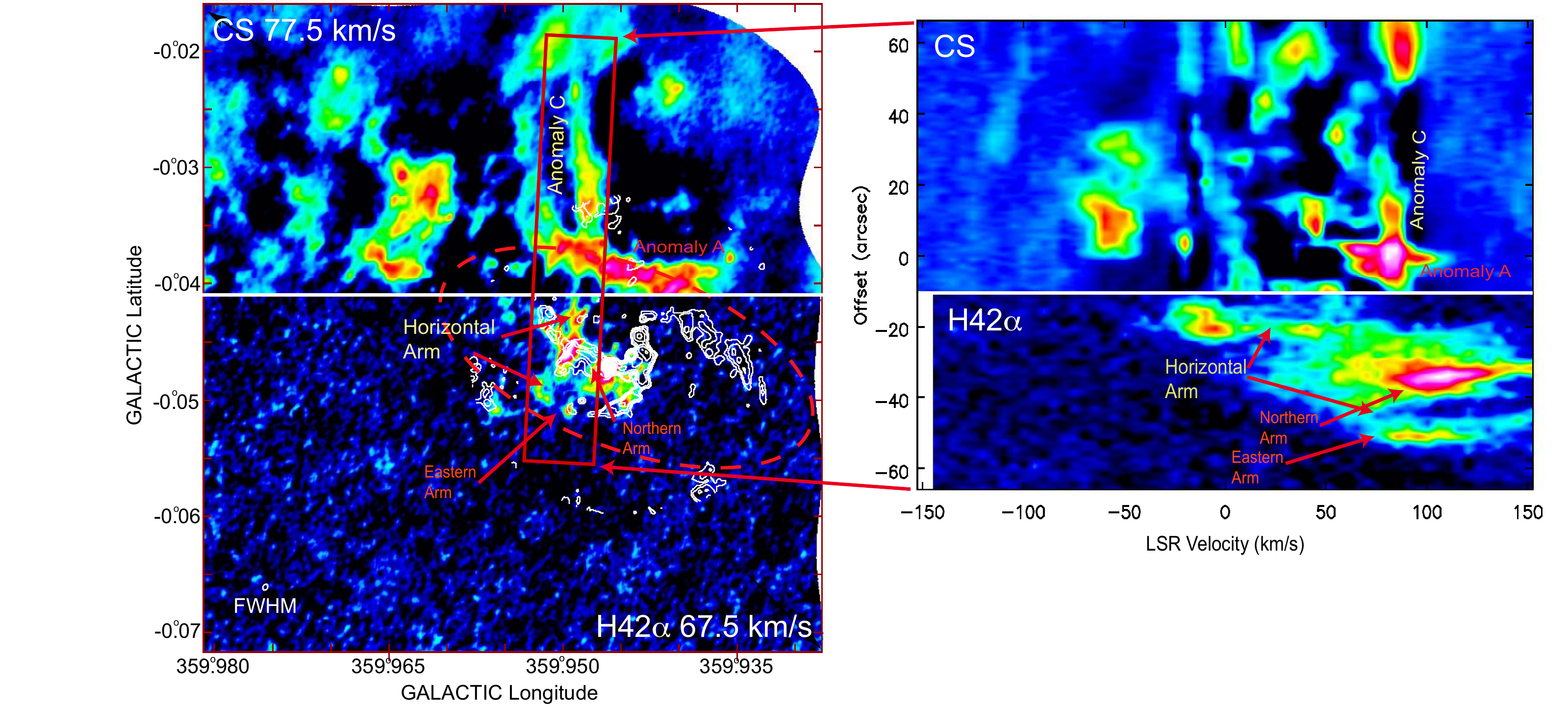}
\end{center}
\caption{Left panel: Integrated intensity map of Anomaly C at $V_{\mathrm{LSR}}=77.5$ km s$^{-1}$ in the CS $J=2-1$ emission line (upper part) and Horizontal Arm at $V_{\mathrm{LSR}}=67.5$ km s$^{-1}$ in the  H42$\alpha$ recombination line (lower part). The central velocities are shown at the upper left and lower right corners of the upper and lower parts, respectively. The velocity widths of the two parts are 10 km s$^{-1}$.  The red dashed ellipse shows the elliptical outline of the CND.   The white contours show the continuum emission at 100 GHz of the ``Galactic Center Mini-spiral" at 100 GHz for comparison \citep{Tsuboi2016}. Upper right panel: PV diagram along Anomaly C in the CS $J = 2 - 1$ emission line. The sampling area is shown by the upper half of the red rectangle in the left panel.
Lower right panel: PV diagram along Horizontal Arm in the H42$\alpha$ recombination line. The sampling area is shown by the lower half of the red rectangle in the left panel.}
\end{figure}

Figure 11 shows the positional relation between the molecular gas of AB in the CS emission line and the ionized gas of Eastern Arm (EA) in the H42$\alpha$ recombination line.  The ionized gas motion in EA has been explained as a part of a Keplarian elliptical orbit which is determined by the appearance and kinematics(e.g. \cite{Zhao2009, Zhao2010,  Tsuboi2017}).   AB and EA are identified as a continuous and united body although they are made of the different gas components, 

Because a compact component is seen  in the SiO emission line around the boundary between AB and EA (see Figure 4),  the molecular gas may have been ionized by strong UV emission from the Central Cluster and  compressed  by the expanding ionized gas.  The area is also identified by SOFIA as a compact component in extremely high-$J$ CO emission lines (up to $J=16-15$) \citep{Requena-Torres12}.
On the other hand, the other area of AB is not clear in the SiO emission line.  
As mentioned in the subsection 3.2.2, AB has filamentary substructures.
The interactions between the filaments may produce mildly shocked molecular gas ($\Delta\sim10$ km s$^{-1}$). 
Therefore, almost all part of AB is not abundant in strongly shocked molecular gas except for the boundary area. 
Meanwhile the northern half of AB can be also identified in the channel maps with $V_{\mathrm{c, LSR}}=77.5$  to $97.5$ km s$^{-1}$ in the CH$_3$OH emission line. The northern half of AB is abundant in mildly shocked molecular gas.  Therefore the shock velocity  in the structure is probably in the range of 10 to 30 km s$^{-1}$. 
The southern half of AB cannot be identified in the channel maps of $V_{\mathrm{c, LSR}}=107.5$ and $117.5$ km s$^{-1}$  except for the small patch located around $l\sim359.960^\circ, b\sim-0.034^\circ$ (see Figure 6).  
Consequently, we argue that AB is falling from the outer region to the vicinity of Sgr A$^\star$, being disrupted by the tidal shear of Sgr A$^\star$ and affected by cosmic ray. 
When AB is further approaching to Sgr A$^\star$, it would be immediately ionized by extremely strong Lyman continuum emission ($\sim4\times10^{50}$ s$^{-1}$) from the Central Cluster (e.g. \cite{Scoville}). The final destination of the falling clouds depends on the initial conditions including impact parameter as mentioned above. Because  AB is ionized and connects with EA, the impact parameter of AB is considered to be smaller than that of AA. 

Figure 12 shows the positional relation between the molecular gas of AC  in the CS emission line and the ionized gas of Horizontal Arm (HA) in the H42$\alpha$ recombination line. HA is identified as a nearly vertical substructure of the ionized gas in the Galactic coordinates.  AC and HA are apparently identified as a continuous and nearly vertical filament  in the channel map.  However, the AC is not continuous to HA in the position velocity diagram.
This would be also a falling  molecular gas from the outer region to the vicinity of Sgr A$^\star$ but is not ionized. 
The impact parameter of AC seems larger than that of AB. 

\section{Conclusions}
We present  the high angular resolution and high sensitivity images  of the  CND and the surrounding region in the CS $J=2-1$, SiO $v=0~J=2-1$, H$^{13}$CO$^+ J=1-0$, C$^{34}$S $J=2-1$, and CH$_3$OH $J_{K_a, K_c}=2_{1,1}-1_{1,0}A_{--}$  emission lines using ALMA. 
\begin{itemize}
  \item The CND is depicted as a torus-like molecular gas with the inner radius of $R_{\mathrm{in}}\sim1.5$ pc and the outer radius of $R_{\mathrm{out}}\sim2$ pc in these emission lines except for the  CH$_3$OH emission line.
  \item The kinematics of the CND roughly obeys rotation with some radial motion. The rotation velocity and radial motion of the CND are estimated to be $V_{\mathrm{rot}}\sim115$ km s$^{-1}$   and $V_\mathrm{rad}\sim23$ km s$^{-1}$, respectively. 
  
  \item The LTE molecular gas mass of the CND  is estimated to be $M_{\mathrm{LTE}}\sim3.1\times10^4 M_\odot$ from the observations of the CS $J=2-1$ and C$^{34}$S $J=2-1$ emission lines.
\end{itemize} 
Our derived  size, kinematics, and mass of the CND are consistent with those in the previous observations.  

We also found some anomalous molecular clouds in the surrounding region, which have filamentary appearances prominent in the CS $J=2-1$ emission line.  
\begin{itemize}
  \item AA is positionally connected with the northwestern sector of the CND, which is adjacent to WA. This is likely being disrupted by the tidal shear of Sgr A$^\star$. The cloud seems to rotate around Sgr A$^\star$ in the opposite direction to the CND.  The molecular cloud would be recently  falling from the outer region to the CND because  the relaxation is not yet occurred.
 \item AB is continuously connected with the eastern tip of the ``Eastern Arm (EA)" of the GCMS. The velocity of the cloud is consistent with that of the ionized gas in the EA. These facts suggest that the molecular cloud is falling from the outer region to  the vicinity of Sgr A$^\star$, being disrupted by the tidal shear, and ionized by strong UV emission from the Central Cluster because the impact parameter of the cloud is smaller than that of AA.
 \end{itemize}
 These clouds would play an important role in  transferring material from the outer region of the Sgr A complex to the CND and/or the vicinity of Sgr A$^\star$.

\begin{ack}  
This work is supported in part by the Grant-in-Aid from the Ministry of Education, Sports, Science and Technology (MEXT) of Japan, No.16K05308.
This work makes use of the following ALMA data: ADS/JAO.ALMA\#2012.1.00080.S 
and ALMA\#2012.1.00543.S.  The later data are retrieved from the JVO portal (http://jvo.nao.ac.jp/portal) operated by the National Astronomical Observatory of Japan (NAOJ).  The National Radio Astronomy Observatory (NRAO) is a  facility of the National Science Foundation operated under cooperative  agreement by Associated Universities, Inc. ALMA is a partnership of ESO (representing its member states), NSF (USA) and NINS (Japan), together with NRC (Canada), NSC and ASIAA (Taiwan), and KASI (Republic of Korea), in cooperation with the Republic of Chile. The Joint ALMA Observatory is operated by ESO, AUI/NRAO and NAOJ. This work has made use of NASA's Astrophysics Data System.
\end{ack}

\clearpage

\end{document}